\documentclass[sigconf]{acmart}

\usepackage{algorithm,algpseudocode}

\usepackage{multirow, makecell}
\usepackage{subcaption}
\usepackage{booktabs}
\usepackage{enumitem}
\usepackage[skip=2pt]{caption}

\usepackage{dingbat}
\usepackage{pifont}
\usepackage{amsmath}

\usepackage{fontawesome}

\usepackage{caption}
\newcommand{\ra}[1]{\renewcommand{\arraystretch}{#1}}
\newcommand{\cmarknew}{\ding{51}}%
\newcommand{\xmarknew}{\ding{55}}%
\setcopyright{none} % No copyright notice required for submissions
\acmConference[Anonymous Submission to The Web Conference]{The Web Conference}
\acmYear{2022}

\newcommand{\model}[0] {\textit{Hetero-SCAN}}

\begin{document}
\title{Meta-Path-based Fake News Detection Leveraging Multi-level Social Context Information} % TODO: replace with your title
\author{Jian Cui}
\email{cj19960819@kaist.ac.kr}
\affiliation{%
  \institution{Korea Advanced Institute of Science and Technology}
  \state{Daejeon}
  \country{South Korea}
}

\author{Kwanwoo Kim}
\email{kw2128@kaist.ac.kr}
\affiliation{%
  \institution{Korea Advanced Institute of Science and Technology}
  \state{Daejeon}
  \country{South Korea}
}

\author{Seung Ho Na}
\email{harry.na@kaist.ac.kr}
\affiliation{%
  \institution{Korea Advanced Institute of Science and Technology}
  \state{Daejeon}
  \country{South Korea}
}

\author{Seungwon Shin}
\email{claude@kaist.ac.kr}
\affiliation{%
  \institution{Korea Advanced Institute of Science and Technology}
  \state{Daejeon}
  \country{South Korea}
}

\begin{abstract}

Fake news, false or misleading information presented as news, has a significant impact on many aspects of society, such as in politics or healthcare domains. Due to the deceiving nature of fake news, applying Natural Language Processing (NLP) techniques to the news content alone is insufficient. Therefore, more information is required to improve fake news detection, such as the multi-level social context (news publishers and engaged users in social media) information and the temporal information of user engagement. The proper usage of this information, however, introduces three chronic difficulties: 1) multi-level social context information is hard to be used without information loss, 2) temporal information is hard to be used along with multi-level social context information, 3) news representation with multi-level social context and temporal information is hard to be learned in an end-to-end manner. To overcome all three difficulties, we propose a novel fake news detection framework, \model{}. We use Meta-Path to extract meaningful multi-level social context information without loss. Meta-Path, a composite relation connecting two node types, is proposed to capture the semantics in the heterogeneous graph. We then propose Meta-Path instance encoding and aggregation methods to capture the temporal information of user engagement and learn news representation end-to-end. According to our experiment, \model{} yields significant performance improvement over state-of-the-art fake news detection methods.

% Moreover, we use knowledge graph embedding methods and Recurrent Neural Network (RNN) to include the temporal information of user engagement in the news representation.
% We use Meta%Considering that even people cannot easily distinguish fake news by news content, applying Natural Language Processing (NLP) techniques on the news content is insufficient.
% To handle this emerging problem, many fake news detection methods have been proposed, applying Natural Language Processing (NLP) techniques on the news content.
%To fully utilize these information, we need to overcome three main challenges: 1) the multi-level social context information should be used without information loss, 2) the news representation should be learned in an end-to-end manner, 3) the temporal information of user engagement should be used. 
% Moreover, we use knowledge graph embedding methods to encode node features extracted from Meta-Path and feed encoded vectors into the Recurrent Neural Network (RNN) to include the temporal information in the news representation.
% However, it is challenging to incorporate those multi-level social context information to detect fake news since those information is highly heterogeneous.
% In order to efficiently incorporate those information in fake news detection, we adopt the concept, Meta-Path, which is a sequence of node types connected with different relations.
% Meta-Path enables us to include such rich social context information in a semantically meaningful way. 
% By using the Meta-Path, we propose an end-to-end fake news detection framework, \model{}.

\end{abstract}

% TODO: replace this section with code generated by the tool at https://dl.acm.org/ccs.cfm



% -- end of section to replace with generated code

% TODO: replace with your keywords

\maketitle

\section{Introduction}

The wide dissemination of fake news has become a major social problem in the world.
The most recent and infamous distribution of fake news was in the 2020 United States presidential election fraud~\cite{election_fraud} and COVID-19 rumors~\cite{covid_rumours}.
Both industry and government are making efforts to prevent the spread of fake news~\cite{fb_report}.
Nevertheless, fake news verification still relies on human experts and their manual efforts in analyzing the news contents with additional evidence.
Therefore, there should be an automatic and efficient way to identify the veracity of the news.

The most typical way to detect fake news is applying Natural Language Processing (NLP) techniques on the news content~\cite{conroy2015automatic, feng2012syntactic}.
Considering that even people struggle in identifying the news authenticity by the news content alone, these NLP solutions are ineffective.
Thus, more information is required to improve fake news detection.

\begin{figure}[t]
    \centering
    \includegraphics[width=\linewidth]{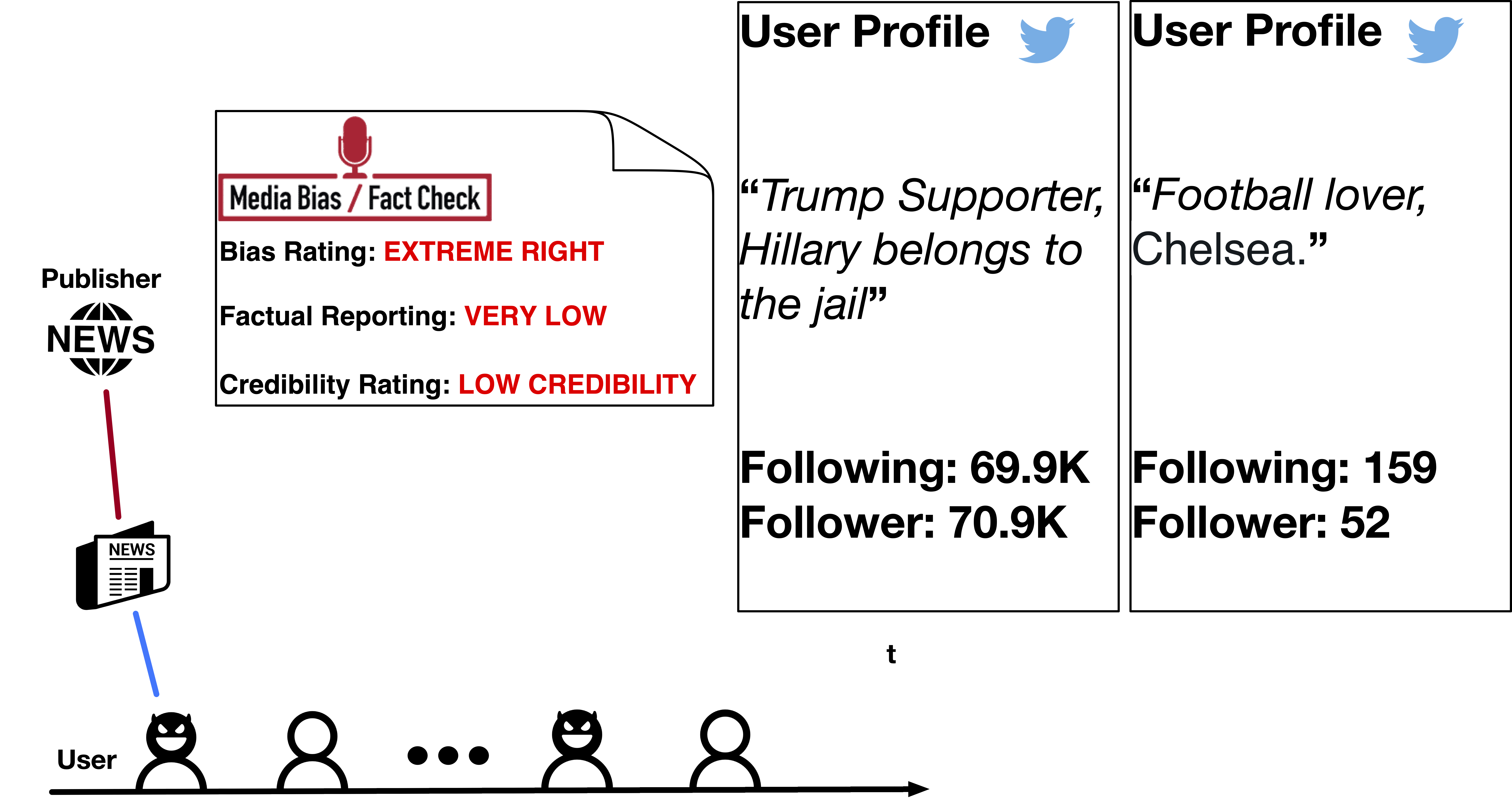}
    \caption{Example of fake news distribution and dissemination. Publishers publish the news, and users tweet the news. Some publishers are regarded as low credibility sources according to the famous fact-checking website, MBFC. User A is an example of an instigator in Twitter, and User B is an example of a regular user.}
    \label{fig:motivation}
    \vspace{-17pt}
\end{figure}

The first important information is the users in social media.
Social media is one of the most influential mediums to propagate information, and it has become a common practice for people to share their thoughts in social media.
Even though regular users use social media as a communication tool, some users, known as instigators, intentionally spread fake news.
Instigators usually have a highly partisan-biased personal description and a lot of followers and followings, which is significantly different from the profiles of regular users (See in Figure~\ref{fig:motivation}).
Therefore, analyzing the users engaged in the news can provide additional evidence for identifying news authenticity.
The publisher information can also play an important role because certain partisan-biased publishers are more likely to publish fake news~\cite{mbfc_right, mbfc_left, mbfc_source}.
Information on users and publishers can be viewed as multi-level social context information, and they provide additional clues for fake news detection.
% \raggedbottom

In addition to multi-level social context information, temporal information of user engagement (temporal information for short) is another instrumental information in fake news detection.
Fake and real news show different propagation properties in social media: Fake news is periodically mentioned by people and usually lasts longer, but real news receives attention only at the beginning of the news publication~\cite{kwon2013prominent}.
In this context, the temporal information should be included in the news representation along with multi-level social context information.

Using multi-level social context and temporal information, however, leads to three chronic difficulties.
Firstly, due to the heterogeneity of multi-level social context information, it is hard to use this information without loss.
Secondly, temporal information is hard to be used along with multi-level social context information.
The graph is a typical way to present social context and its connectivity to the news, but the graph itself has complications in presenting temporal information.
The last difficulty is to learn the news representation end-to-end.
Multi-level social context and temporal information are two different kinds of information, which increases the difficulty of adopting end-to-end learning while utilizing both information.
To promise a high-performing fake news detection, it is necessary to adopt end-to-end learning. It enables us to eliminate the effect from the sub-tasks and optimize the training parameters with a single news detection objective.
% To sum up, the goal of this paper is to learn the news representation end-to-end by utilizing both multi-level social context and temporal information without loss.

To the best of our knowledge, existing approaches fail to address all three difficulties, so we propose a novel fake news detection framework, \model{}, to tackle above-listed difficulties.
In \model{}, to preserve multi-level social context information, we use the \emph{Meta-Path}.
Meta-Path is a composite relation connecting two node types, aiming to capture the semantics in the heterogeneous graph.
%% [Definitions from other paper]
% Meta-path is a path defined on the network schema, which is a relation sequence between two object types and defines a new or existing relation between objects.
We define two Meta-Paths containing different aspects of news (users and publishers) to extract multi-level social context information without information loss.
Moreover, Meta-Path instance encoding and aggregation methods are proposed to capture the temporal information of user engagement and learn the news representation end-to-end.

To show that our proposed method outperforms existing solutions, we test \model{} with two real-world datasets~\cite{nguyen2020fang, dai2020ginger}, and the results show that \model{} achieves significant improvement over previous approaches in terms of F1 score, accuracy, and AUC score. 
Our code with data is released on the GitHub~\footnote{https://github.com/(anonymous)/hetero\_scan} for reproducibility. 
Our major contributions are:
\begin{enumerate}[leftmargin=.3in, topsep=3pt]
    \item We pose three chronic difficulties in social context aware fake news detection and address them by proposing a novel fake news detection framework, \model{}.
    \item We conduct diverse experiments on the two real-world fake news datasets, covering the broad definition of fake news (Section~\ref{sec:preliminaries}), and demonstrate that \model{} shows better performance than existing solutions.
    \item We provide new insights into the differences in the behavior of engaged users between intentional and unintentional fake news.
\end{enumerate}

% A general approach to present multi-level social context is using a heterogeneous graph, where nodes are publishers, news, and users.
% To fully utilize multi-level social context information in the graph and thus improve the detection performance, three chronic difficulties need to be resolved.
% we pose three crucial yet overlooked challenges.

\section{Related Work}

\subsection{Fake News Detection}
Fake news detection methods can be categorized into two types: content-based and graph-based approaches. 

The content-based approach models the content of the news, such as headline or body text, to detect news authenticity. 
Some research on content-based approaches utilizes linguistic features such as stylometry, psycholinguistic properties, and rhetorical relations~\cite{potthast2018stylometric, perez2018automatic, rubin2015truth, castillo2011information}.
%They believe that there are different features in between real and fake news.
Researchers also use Multi-modal approaches, the combination of visual and linguistic features to verify the news authenticity~\cite{khattar2019mvae,giachanou2020multimodal, wang2018eann, vo2021hierarchical, qian2021hierarchical}.
%Some other approaches employ knowledge as a feature~\cite{ciampaglia2015computational,shi2016discriminative},to compare knowledge triples from the news with existing reliable sources; if a certain fraction of the knowledge triples correspond with reliable knowledge, then the news is considered to be a real.

The graph-based approach, also known as the social context aware approach, adds auxiliary information of the user or publisher to model the news.
% Previously, researchers utilized RNN to deal with time series of social engagement~\cite{ruchansky2017csi, ma2016detecting}. 
% More complex approaches appear to introduce graph features.
% Tri-FN~\cite{shu2019beyond} uses the matrix decomposition technique to obtain the latent feature of the news with social context.
% Using matrix decomposition has a high computational cost, so it limits its applicability in the real world.
CSI~\cite{ruchansky2017csi} is a framework that aims to capture the information of users and their temporal engagements.
CSI, however, does not consider publishers, and the connection between users and news was also ignored.
Bi-GCN~\cite{bian2020rumor} and SAFER~\cite{chandra2020graph} use Graph Convolution Network (GCN)~\cite{kipf2016semi} to obtain the news representation with user information.
However, they suffer from a severe information loss since they present news and user information in a homogeneous graph. 
In other words, they fail to taking the node and relation types into account.
Most recently, FANG~\cite{nguyen2020fang} is proposed to preserve information by dividing the fake news detection task into several sub-tasks, such as textual encoding and stance detection.
Nonetheless, dividing into sub-tasks causes the error propagation problem: If the sub-tasks have errors, the errors can propagate up to the final news representation and thereby deteriorate the detection performance.
AA-HGNN~\cite{ren2020adversarial} uses adversarial active learning and extends Graph Attention Network (GAT)~\cite{velivckovic2017graph} into the heterogeneous graph to learn the news representation with limited training data.
Information of users and their temporal engagement information, however, are not considered in AA-HGNN.
Table~\ref{t:related_work} compares \model{} and existing fake news detection methods.

% To efficiently detect fake news, there are several e
% However, these existing graph-based methods still suffer from the following four major drawbacks: 1) expensive computational cost due to a large number of user nodes in the graph, 2) the error in sub-tasks, such as textual encoding or stance detection propagates to the final news detection loss, 3) information loss caused by ignoring the node types and relations in between, and 4) temporal information is missing in the final news representation. 
% In our work, we build a heterogeneous graph with social users and publishers to capture their information, including temporal.
% This rich information is modeled by our proposed model efficiently in an end-to-end manner. 
% By doing so, our method overcomes the shortcomings of the previous research. 

\begin{table}[t]
    \centering
    \caption{Comparison of \model{} with exiting graph-based fake news detection methods.}
    \ra{1.2}
	\begin{tabular}{ccccc}
	\toprule
    % & \multirowcell{2}{\small Expensive \\ \small computation} & \multirowcell{2}{ \small Information \\ \small loss} & \multirowcell{2}{\small Error \\ \small propagation} & \multirowcell{2}{\small Missing \\ \small temporal} \\
    & \multirowcell{2}{\small Multi-level \\ \small Social Context} & \multirowcell{2}{\small Information \\ \small Preserving} & \multirowcell{2}{\small Temporal \\ \small Information} & \multirowcell{2}{\small End-to\\ \small -end}
    \\	\\
	\midrule
	\footnotesize CSI~\cite{ruchansky2017csi}  & \xmarknew & \cmarknew & \cmarknew  & \cmarknew\\
	\footnotesize SAFER~\cite{chandra2020graph}   & \xmarknew & \xmarknew & \xmarknew  & \cmarknew\\
	\footnotesize FANG~\cite{nguyen2020fang}    & \cmarknew & \cmarknew & \cmarknew  & \xmarknew\\
	\footnotesize AA-HGNN~\cite{ren2020adversarial} & \xmarknew & \cmarknew & \xmarknew  & \cmarknew\\
	\midrule
	\small \model{} & \cmarknew & \cmarknew & \cmarknew & \cmarknew \\
	\bottomrule
    \end{tabular}
    \label{t:related_work}
    \vspace{-15pt}
\end{table}

\subsection{Graph Neural Network} 
Graph Neural Network, the extension of the deep learning method into graphs, shows its effectiveness in graph-represented data. 
The first method proposed is Graph Convolutional Network (GCN)~\cite{kipf2016semi} which aggregates the features from the adjacent nodes in the graph. 
To further improve it, some methods adopt the attention mechanism and random work with restart sampling strategy, namely Graph Attention Network (GAT)~\cite{velivckovic2017graph} and GraphSAGE~\cite{hamilton2017inductive}.

As these methods are designed for homogeneous graphs, they are not general enough to apply to the heterogeneous graph, so new approaches tailor to heterogeneous graphs are then proposed.
To model the multi-relations in the graph, the Relation aware GCN (R-GCN)~\cite{schlichtkrull2018modeling} is proposed first.
HetGNN~\cite{zhang2019heterogeneous} uses a sampling strategy based on random walk with restart and Bi-LSTM to aggregate the node features in the heterogeneous graph. 
Later, the methods based on Meta-Path and attention mechanism, such as HAN~\cite{han2020graph} and MAGNN~\cite{fu2020magnn}, are proposed.
% Though there are some works proposing relation aware GCN~\cite{schlichtkrull2018modeling}, they are not general enough to apply on a heterogeneous graph. 
% Though these method can be directly applied to the heterogeneous graph of news, it did not give an idea re
% Similar to HAN and MAGNN, \model{} is based on Meta-Path extraction, but includes temporal awareness, capturing the propagation property of news dissemination. 
% The performance of \model{} outperforms the previous graph-based fake news detection approaches and the GNN methods. 
\section{Preliminaries} \label{sec:preliminaries}

\begin{definition}[\textbf{Broad Definition of Fake News}]
Fake news is false news.
\end{definition}

\begin{definition}[\textbf{Narrow Definition of Fake News}]
Fake news is intentionally false news published by a news outlet.
\end{definition}

Contrary to the amount of research done, the term fake news has only just been defined by the recent work of Zhou, Xinyi and Reza Zafarani~\cite{zhou2020survey}.
They define fake news in two scopes, broad and narrow. 
The broad definition emphasizes the authenticity of the information, and the narrow one emphasizes the intentions of the author.
Most research on fake news detection has employed a broad definition of fake news.
We experiment on the two dataset (with and without intention) following broad definition and analyze how intention affect the performance of the detection (in Section~\ref{sec:mis_dis}).

\begin{definition}[\textbf{Heterogeneous Graph}]
A heterogeneous graph is defined as a graph $\mathcal{G} = (\mathcal{V}, \mathcal{E})$ associated with a node type mapping function $\phi: \mathcal{V} \rightarrow \mathcal{A}$  and an edge type mapping function $\psi: \mathcal{E} \rightarrow \mathcal{R} $. $\mathcal{A}$ and $\mathcal{R}$ denotes the predefined sets of node types and edge types, respectively, with $|\mathcal{A}| + |\mathcal{R}| > 2$.
\end{definition}

\begin{definition}[\textbf{Meta-Path}]
\label{def:Meta-Path}
A Meta-Path $P$ is defined as a path in the form of ${A_1} \xrightarrow{R_1} {A_2} \xrightarrow{R_2} ... \xrightarrow{R_l} {A_l} $ (abbreviated as ${A_1}{A_2}...{A_l}$), which describes a composite relation $R = {R_1} \circ {R_2} \circ ... \circ {R_n}$ between node types ${A_l}$ and ${A_{l+1}}$, where $\circ$ denotes the composition operator on relations.
\end{definition}

\begin{definition}[\textbf{Meta-Path Instance}]
Given a Meta-Path $P$ of a heterogeneous graph, a Meta-Path instance $p$ of $P$ is defined as a node sequence in the graph following the schema defined by $P$.
\end{definition}

% \begin{definition}[\textbf{Meta-Path based Neighbor}]
% Given a Meta-Path $P$ of a heterogeneous graph, the Meta-Path based neighbors $N_v^P$ of a node $v$ is defined as the set of nodes that connect with node $v$ via Meta-Path instances of $P$. 
% \end{definition}

\section{Methodology}

\subsection{Graph Construction \& Feature Engineering} \label{sec:feat_eng}

% \begin{figure}[h]
%     \includegraphics[width=\columnwidth]{figure/het_graph.pdf} 
%     \caption{Heterogeneous Graph of News}
%     \label{fig:graph}
% \end{figure}

\begin{figure}[t]
    \centering
    \includegraphics[width=\columnwidth]{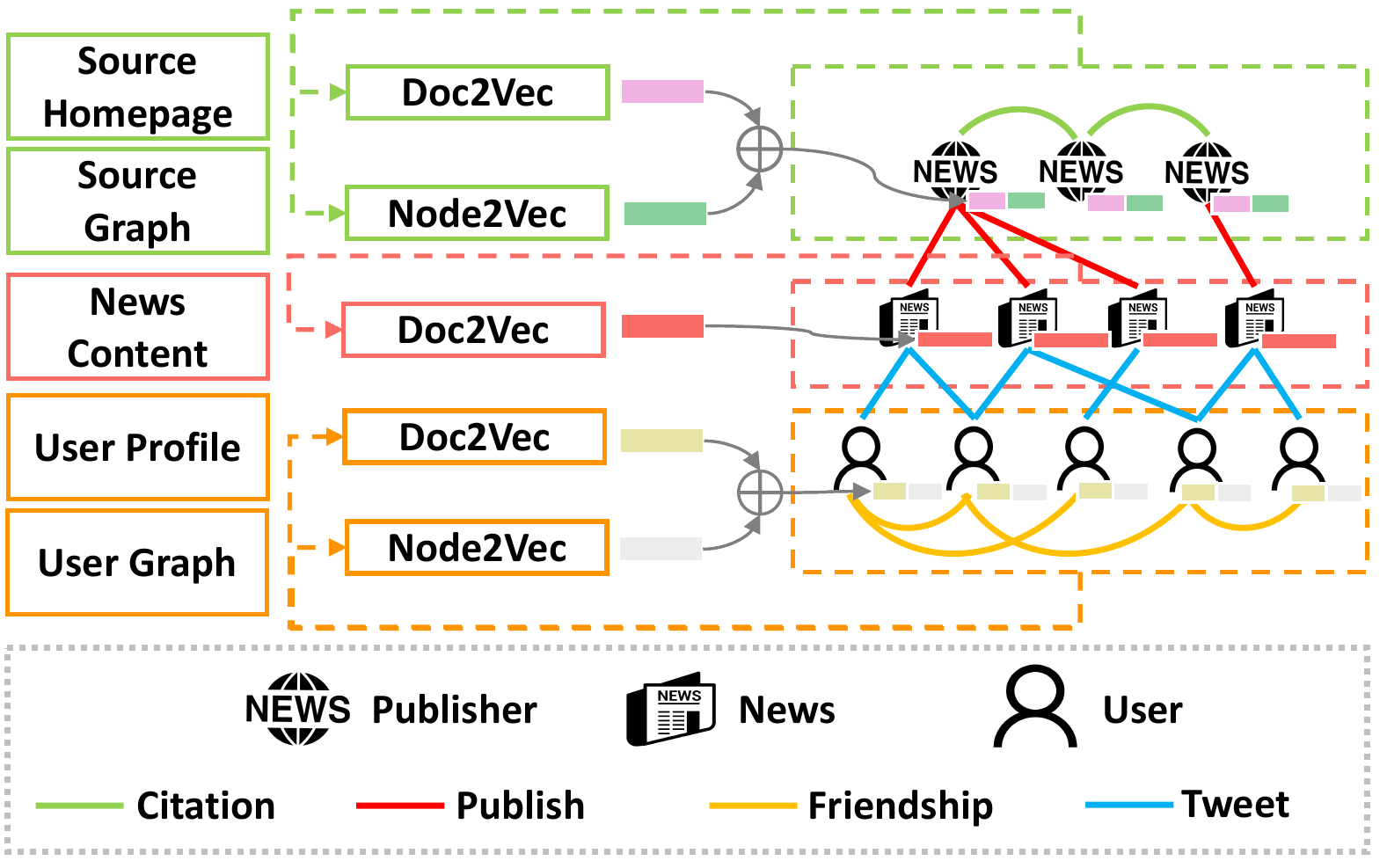} 
    \caption{Heterogeneous Graph of News and Node Feature Engineering.}
    \label{fig:feat_eng}
    \vspace{-15pt}
\end{figure}

% Because considering news content alone is not enough to detect fake news, the user and publisher information is also required for fake news detection.
To integrate multi-level social context information, we build a heterogeneous graph of news (Figure~\ref{fig:feat_eng}).
The graph consists of three types of nodes (publisher, news, and users) and four types of edges (citation, publication, tweet, and following).
Formally, the heterogeneous graph of news is noted as $\mathcal{G}(\mathcal{V},\mathcal{E})$, and the set of three node types are symbolized as $\mathcal{A} = \{A_p, A_n, A_u\}$.

Before utilizing this heterogeneous graph, it is necessary to construct initial node features for three types of nodes in the graph.
For news nodes, Doc2Vec~\cite{le2014distributed} is applied to the news article to construct their initial features.
The user and publisher nodes, however, need additional information to construct their respective initial features.
Users' profiles are used for user nodes since the importance of the user profiles for detecting news authenticity has been proved by Shu, Kai et al.~\cite{shu2019role}.
The distinct feature of each publisher is acquired from about-us pages on their official websites; If there is no about-us page on the publisher's official website, we use Wikipedia's description instead.
Doc2Vec is applied again to leverage these text contents.
To also include the structural role they play in their respective networks, we apply Node2Vec~\cite{grover2016node2vec} to capture user connections and citations among publishers as features.
%\jian{The connection among users and the citation among publishers is another instrumental information, so we apply another unsupervised method, Node2Vec~\cite{grover2016node2vec} to include the structural role of user and publisher in their network.}
By concatenating the two vectors obtained from Doc2Vec and Node2Vec, we construct the initial features of user and publisher nodes.
%We construct initial features of user and publisher nodes by concatenating the twovectors obtained from Doc2Vec and Node2Vec.
Figure~\ref{fig:feat_eng} shows the overall node feature construction process.
% Our Meta-Path based heterogeneous graph embedding method does not yet consider the relationship between the same type of node, i.e., the $User -User$ and $Publisher - Publisher$ relationships.
%% [Description of Node2Vec]
% The Node2Vec is an algorithmic framework for representational learning on graphs. 
% With the help of Node2Vec, we obtain the structural representation of each node in the network they belong to.

\subsection{Meta-Path Instance Extraction} \label{sec:extraction}
After constructing initial node features, we then need to learn the news representation containing multi-level social context and temporal information.
Multi-level social context information should be used without loss, which is the first difficulty in social context aware fake news detection.
To address this difficulty, we use the concept, Meta-Path (defined in Section~\ref{sec:preliminaries}).
Meta-Paths can be used to extract meaningful social context with respect to publishers and users.
We define two Meta-Paths that reflect the method used for actual news verification.
When people verify the news authenticity, they need to cross-check both publisher and the news published by this publisher.
% \jian{Defining meaningful Meta-Paths and extracting Meta-Path instances that follow Meta-Paths are essential for our task.}
% \jian{The publisher information and news published by this publisher need to be reviewed together when identifying the news authenticity. 
The same goes for users: User information, as well as the news tweet by the user, needs to be reviewed.
From these two intuitions, a set of Meta-Path $\mathcal{P}$ that we find useful is defined as below:
\begin{equation}
    \mathbf{\mathcal{P}} \in \{\mathcal{P}_U, \mathcal{P}_S\}
\end{equation}
where $\mathcal{P}_U: News \rightarrow User \rightarrow News$ and $\mathcal{P}_S: News \rightarrow Publisher \\ \rightarrow News$.
% where $\mathcal{P}_U: News \rightarrow User \rightarrow News$ means \textit{"the two news are tweeted by the same user"}, and $\mathcal{P}_S: News \rightarrow Publisher \rightarrow News$ means \textit{"the two news are published by the same publisher"}.

\begin{figure}[t]
    \centering
    \includegraphics[width=\columnwidth]{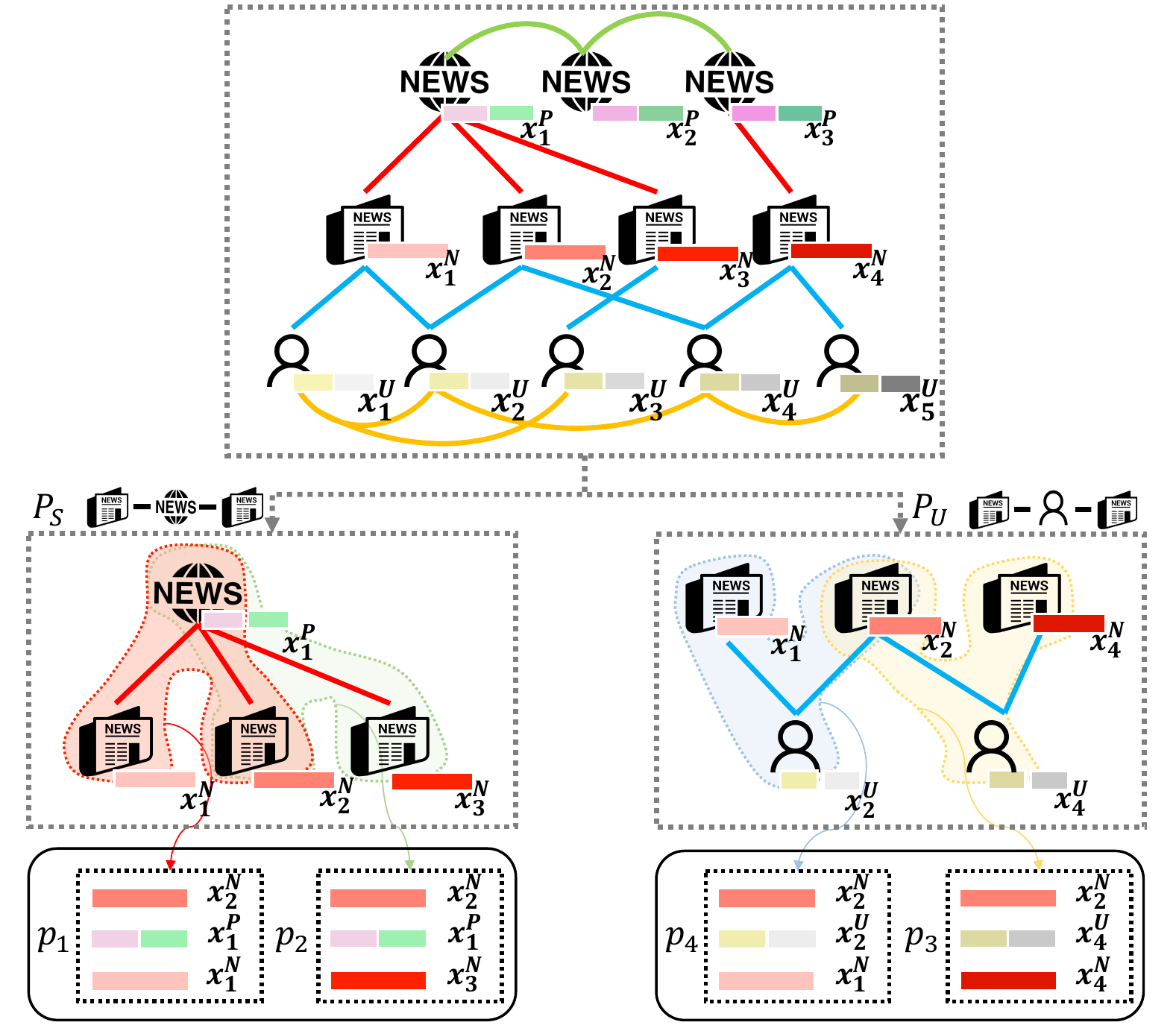}
    \caption{Extracting Meta-Path instances of the target news node $x_2^{N}$.}
    \label{fig:Meta-Path_extraction}
    \vspace{-15pt}
\end{figure}

After defining a set of Meta-Path, we extract Meta-Path instances $p$ following each Meta-Path, $\mathcal{P}_S$ or $\mathcal{P}_U$, for each target news node. 
To efficiently extract Meta-Path instances, we first divide the whole graph into two sub-graphs, which only contain the nodes types specified in the Meta-Path, $\mathcal{P}_S$ or $\mathcal{P}_U$. 
Then, in each sub-graph, the Meta-Path instances following each Meta-Path are extracted. 
The corresponding collection of features are fed into \model{} to get the final representation of the target news node. 
The sets of instances following two Meta-Path  $\mathcal{P}_S$ and $\mathcal{P}_U$ are denoted as $\mathbf{P}_S$ and $\mathbf{P}_U$ respectively.
For instance, if we want to extract the Meta-Path instances of the target news node $x_2^{N}$ in Figure~\ref{fig:Meta-Path_extraction}, we first divide the whole graph into two sub-graphs. 
One is composed of news and publisher nodes, and the other is made of news and user nodes. Then, the Meta-Path instances follow Meta-Path $\mathcal{P}_S$ or $\mathcal{P}_U$ are selected from each sub-graph, and the corresponding features of nodes along these Meta-Path instances will be prepared for our model. 
In particular, the Meta-Path instance $p_1$ is made of features of nodes following Meta-Path $\mathcal{P}_S: News \rightarrow Publisher \rightarrow News$, which is $x_1^{N}$, $x_1^{P}$ and $x_2^{N}$ in the graph. 
In the same manner, $p2$, $p3$ and $p4$ are extracted. 
For the target node $v$, we use $\mathbf{P}_S$ and $\mathbf{P}_U$ to denote the set of Meta-Path instances follow each Meta-Path. In this case, $\mathbf{P}_S = \{p1, p2\}$ and $\mathbf{P}_U = \{p3, p4\}$ are set of Meta-Path instances of target node $x_2^{N}$.

There are usually a large number of users engaged per news in the real world. 
To cope with this situation, we extract Meta-Path instances from our heterogeneous graph of news with random sampling.
Specifically, a certain number of Meta-Path instances are randomly sampled for each news node according to a pre-defined Meta-Path. 
At last, in order to capture the temporal information, the model should be aware of the chronological information of Meta-Path instances. 
Thus, the Meta-Path instances from the Meta-Path $\mathcal{P}_U$ are sorted chronologically before being fed into the proposed model.
% We include an additional sorting unit to sort the Meta-Path instances.
In the following sections, we assume that the Meta-Path instances from $\mathcal{P}_U$ are sorted in chronological order.

\subsection{Model Architecture}

\begin{figure*}
    \centering
    \includegraphics[scale=0.53]{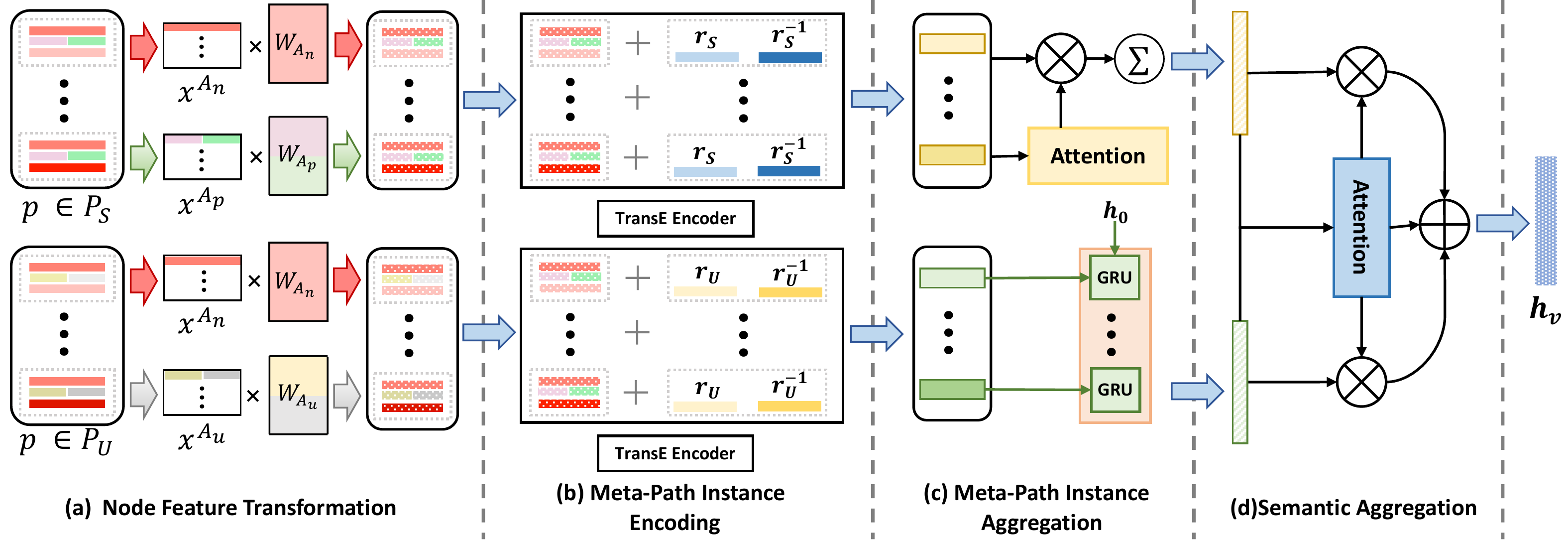} 
    \caption{Architecture of \model{}.}
    \label{fig:model}
    \vspace{-12pt}
\end{figure*}

\model{} takes in vectors from the previous step as input and processes them through four steps as shown in Figure~\ref{fig:model} to tackle the yet addressed chronic difficulties.
%: 1) Node Feature Transformation; 2) Meta-Path Instance Encoding; 3) Meta-Path Instance Aggregation; 4) Semantic Aggregation.

%To tackle two remaining difficulties, the proposed framework, \model{}, takes in the vectors from the previous step as input, and processes them through four steps as shown in Figure~\ref{fig:model}: 1) Node Feature Transformation; 2) Meta-Path Instance Encoding; 3) Meta-Path Instance Aggregation; 4) Semantic Aggregation.

\subsubsection{\textbf{Node Feature Transformation}} \label{sec:feat_transform}
The initial node features have different dimensions since different sources and techniques are used in the feature engineering process (Section~\ref{sec:feat_eng}).
To make them lie in the same latent space, we apply the type-specific linear transform on the features of each type of node.
Type-specific transformation refers to the linear projection of a vector into another dimension for each type of node in the graph.
The transformed feature for a node $v \in \mathcal{V}_{A}$ of type $A \in \mathcal{A}$ is: 
\begin{equation}
\mathbf{h}^{A}_{v} = \mathbf{W}_A \cdot \mathbf{x}_{v}^{A}
\end{equation}
where $\mathbf{x}_v \in \mathbb{R}^{d_A}$ is the initial feature of node $v$, and $\mathbf{W}_A \in \mathbb{R}^{d^{'} \times d_A}$ is the learnable type-specific weight matrix for node type $A$.

\subsubsection{\textbf{Meta-Path Instance Encoding}} \label{method:encoding}
% After transformation, the input is in the form of three transformed features of nodes that follow either Meta-Path $P_U$ or $P_S$.
The first step transformed all the features of the node into the same dimension.
We then need to efficiently summarize the Meta-Path instances for the remaining aggregation steps, which is important in capturing temporal information and learning the representation end-to-end.
To efficiently encode node features, we adopted the method that shows excellent performance in knowledge graph triple embedding~\cite{wang2014knowledge, sun2019rotate, dettmers2018convolutional}.

The major advantage of using knowledge graph triple embedding is the structural similarity between knowledge graph triples and our Meta-Paths.
In the knowledge graph, the knowledge graph triple usually refers to the subject, predicate, and object $(s, p, o)$. 
The Meta-Path we defined is similar to the knowledge graph triple in a sense that Meta-Path is the same format along with one more entity and relation. 
Formally,
\begin{equation}
\begin{split}
\label{eq:spo_mp}
\text{Knowledge graph triple:} &\mathbf{e}_s \xrightarrow{\mathbf{e}_p} \mathbf{e}_o \\
\text{Meta-Path:} &\mathbf{h}_{u} \xrightarrow{r} \mathbf{h}_w \xrightarrow{r^{-1}} \mathbf{h}_v\\ 
% v \in A_n, & u \in A_n, w \in \{A_p, A_u\} \\
\end{split}
\end{equation}
where $v$ is target node, $u$ and $w$ refer to the nodes along the Meta-Path. 
Considering the Meta-Path we defined in the Section~\ref{sec:extraction}, $v \in A_n$, $u \in A_n$, and $w \in \{A_p, A_u\}$.
The $r$ and $r^{-1}$ is the relation between $u$, $w$ and $w$, $v$ respectively. 
$\mathbf{h}$ is the transformed embedding of the node as we stated in Section~\ref{sec:feat_transform}, and $\mathbf{e}$ is the embedding of the knowledge graph triple.

% To learn the representations of entities and relations in knowledge graphs, many studies have tackled missing link or entity prediction in the past.
Several research on knowledge graph domain tackle the triple embedding problem~\cite{wang2014knowledge, sun2019rotate, dettmers2018convolutional}.
% \textcolor{blue}{[brief information of CONVE and ROTATE] -> done}
% RotatE~\cite{sun2019rotate} maps the entity and relation into the complex space, and it considers relations as rotation.
% ConvE~\cite{dettmers2018convolutional} reshapes embeddings of entity and relations and apply the convolution operation on it. 
We use TransE~\cite{wang2014knowledge} as our main encoding method for the proposed model.
TransE~\cite{wang2014knowledge} represents relations as translations, so the object vector $\mathbf{e}_o$ in the triple is considered as a translation of subject vector $\mathbf{e}_s$ on predicate vector $\mathbf{e}_p$. 
Other than TransE, RotatE~\cite{sun2019rotate} and ConvE~\cite{dettmers2018convolutional} knowledge graph embedding methods are also examined in our work. 
Ablation study on different knowledge graph triple embedding methods and their descriptions are provided in the Appendix.

In knowledge graph, there are usually explicit features for predicated ($\mathbf{e}_p$ in Equation~\ref{eq:spo_mp}), but in our case, there is no explicit features for the relations ($r$ in Equation~\ref{eq:spo_mp}), so we use learnable embedding vectors to present relations.
Inverse relationships, such as $Publisher - News$ and $News - Publisher$, are represented by taking the sign inverses.
For instance, if we define $r$ as the embedding of $Publisher - News$ relationship, the inverse relationship, $News - Publisher$ is $r^{-1} = -r$.
Our encoding function $f_{enc}$ is defined as:
\begin{equation}
    \mathbf{h}_{p} = f_{enc}(p) = f_{enc}(\mathbf{h}_u, r, \mathbf{h}_w, r^{-1})
\end{equation}

\begin{table}[t]\centering
    \caption{Formulation of Encoding Method.}
    \ra{1.5}
 	\begin{tabular}{cll}
	\toprule
	\textbf{Method} & \textbf{Original} & \textbf{In Our Paper} \\ 
	\hline
    TransE & $\mathbf{e}_s+\mathbf{e}_p$ & $MEAN[(\mathbf{h}_u + r + r^{-1}), (\mathbf{h}_{w}+r^{-1})]$ \\
    ConvE  & $[\mathbf{e}_s \mathbin\Vert \mathbf{e}_p] * \mathbf{W}$ &$[\Tilde{\mathbf{h}}_{u} \mathbin\Vert \Tilde{r} \mathbin\Vert \Tilde{\mathbf{h}}_{w} \mathbin\Vert \Tilde{r}^{-1}] * \mathbf{W}$ \\
    RotatE & $\mathbf{e}_s \odot \mathbf{e}_p$ & $MEAN[(\mathbf{h}_u \odot r \odot r^{-1}), (\mathbf{h}_w \odot r^{-1})]$ \\
	\bottomrule
    \end{tabular}
    \label{t:encoder}
    \vspace{-12pt}
\end{table}

The existing knowledge graph triple embedding methods explained above are designed for two nodes and the relation between them.
In our Meta-Path, we have a total of three nodes and two relations in a Meta-Path instance.
We deal with this by slightly tuning the formulation to fulfill our needs. 
The original formulation of knowledge graph triple embedding methods and ours are summarized in Table~\ref{t:encoder}.
In this table, the $\Tilde{\mathbf{h}}$ means the reshape of vector $\mathbf{h}$ in a 2D form, and the $\odot$ and $\mathbin\Vert$ represent the element-wise product and concatenation of vector, respectively.

\subsubsection{\textbf{Meta-Path Instance Aggregation}} \label{method:semantic_aggr}
The encoded vectors from two different Meta-Paths are aggregated by using different methods.

The encoded vectors from Meta-Path $\mathcal{P}_S: News \rightarrow Publisher \rightarrow News$ contain information of other news from the same publisher.
Among the news published by the publisher, not all news will contain valuable information for detection.
Thus, the model should 'focus' on some of the news published by this publisher and include this information in the aggregated representation.
For each Meta-Path instance $p\in\mathbf{P}_S$:
% The encoded Meta-Path instances following $\mathbf{P}_S: News \rightarrow Publisher \rightarrow News$ is aggregated via attention mechanism. 
\begin{equation}
\begin{split}
    e_{p} &= LeakyReLU(\mathbf{a}^{T} \cdot \mathbf{h}_{p}) \\
    \alpha_{p} &= softmax(e_{p}) = \frac{exp(e_{p})}{\sum_{p' \in \mathbf{P}_{S}}exp(e_{p'})} \\
\end{split}
\end{equation}
where $e_{p}$ is the attention value calculated by multiplying encoded Meta-Path instance $\mathbf{h}_{p}$ with attention vector $\mathbf{a} \in \mathbb{R}^{2d^{'}}$, and it is normalized by a softmax function over all Meta-Path instances of the target node $v$, the result is denoted as $\alpha_{p}$ above. 

To alleviate the effect of the high variance of the data in a heterogeneous graph, we adopt multi-head attention mechanism.
$K$ independent attention mechanisms execute the transformation as shown in Equation~\ref{eq:multihead_attn}, and their features are concatenated after they pass the activation function $\sigma$.
The output feature representation can be formulated as:
\begin{equation}
\label{eq:multihead_attn}
\mathbf{h}_v^{\mathcal{P}_S} = \overset{K}{\underset{k=1}{\mathbin\Vert}} \sigma (\sum_{p \in \mathbf{P}_S} [\alpha_{p}]_{k} \cdot \mathbf{h}_{p})
\end{equation}
where $[\alpha_{p}]_{k}$ is the normalized attention value of Meta-Path instance $p$ of target node $v$ at the $k$-th attention head.

Temporal information of user engagement is another critical feature to determine the veracity of the given news, and incorporating this information is the second difficulty to resolve.
To capture the temporal information, we aggregate the Meta-Path instances follow $\mathcal{P}_U: News \rightarrow User \rightarrow News$ through Recurrent Neural Network (RNN). 
Since Meta-Path instances are already encoded in the previous step, we can directly feed them into the RNN.
There are usually a large number of users engaged per news, so we choose GRU~\cite{chung2014empirical} as our RNN unit to avoid the vanishing or exploding gradients problem.
\begin{equation}
\mathbf{h}_v^{\mathcal{P}_U} = \textbf{GRU} (\mathbf{h}_{p_1}, \mathbf{h}_{p_2}, ..., \mathbf{h}_{p_n}), p_i \in \mathbf{P_U}
\end{equation}
The last hidden state of the GRU is used for the downstream task as it is the high-level representation that summarizes the temporal information of the user engagement.

\subsubsection{\textbf{Semantic Aggregation}}
Two vectors, $\mathbf{h}_v^{\mathcal{P}_S}$ and $\mathbf{h}_v^{\mathcal{P}_U}$, from previous step represents two different aspects of the news. 
The final news representation is produced by fusing these two vectors, which enables us to learn the news representation end-to-end (the third difficulty).
As two Meta-Paths show two different aspects of a given news, the model should be able to weigh the importance of the two aspects with different news.
To this end, we adopt another attention mechanism. Before applying the attention mechanism, non-linear transformations are applied to summarize $\mathbf{h}_v^{\mathcal{P}_S}$ and $\mathbf{h}_v^{\mathcal{P}_U}$.
Thus for $P \in \{\mathcal{P}_S, \mathcal{P}_U\}$: 
\begin{equation}
    s_{P} = \frac{1}{|\mathcal{V}|} \sum_{v \in \mathcal{V}} tanh(\mathbf{M}_A \cdot \mathbf{h}_v^{P} + \mathbf{b}_A)
\end{equation}
Here, $\mathbf{M}_A \in \mathbb{R}^{d_{m} \times d^{'}}$ and $\mathbf{b} \in \mathbb{R}^{d_{m}}$ is a learnable weight matrix and bias vector. $\mathcal{V}$  is the set of news nodes.

Then we apply the attention mechanism to aggregate two vectors to obtain our final news representation $\mathbf{h}_v$.
\begin{equation}
\begin{split}
    e_P &= tanh (q^T \cdot s_P) \\
    \beta_P &= \frac{exp(e_P)}{\sum_{P^{'} \in \mathcal{P}}exp(e_{P^{'}})} \\
    \mathbf{h}_v &= \sum_{P \in \mathcal{P}}\beta_p \cdot \mathbf{h}_v^P 
\end{split}
\end{equation}
where $q \in \mathbb{R}^{d_m}$ is the attention vector and $\beta_P$ is the normalized importance of Meta-Path $P$. 

\subsection{Training}
The final representation of the target news vector is passed to the classification layer to get the classification result. 
During training, our predictions and labels are used to calculate the loss, and we update the learnable parameters of the model by using the back-propagation algorithm. 
The loss function used in \model{} is cross-entropy loss, which is: 
\begin{equation}
\mathcal{L} = -\sum y \log \mathbf{P}_{fake} + (1-y) \log  \mathbf{P}_{real}
\end{equation}
The overall all learning algorithm is summarized in Algorithm~\ref{alg:train} (Appendix).

\section{Experimental Result and Analysis} \label{Evaluation}
\begin{table*}[t]\centering
    \caption{Detection result of two real-word dataset: FANG and FakeHealth. Bold numbers denote the best value in average, and underscored numbers denote the smallest variation ($\pm$ stands for 95\% confidence interval).}
    \ra{1.2}
	\begin{tabular}{crccccc}
	\toprule
	\textbf{Dataset} & \textbf{Classification Method} & \textbf{Precision} & \textbf{Recall} & \textbf{F1 Score} & \textbf{Accuracy} & \textbf{AUC Score} \\ 
	\midrule    
	\multirow{6}{*}{FANG} & Classification Layer & \textbf{0.845$\pm$0.052} & \textbf{0.843$\pm$0.054} & \textbf{0.843$\pm$0.053} & \textbf{0.843$\pm$0.054} & 0.839$\pm0.048$\\
                  & Naive Bayes & 0.839$\pm0.053$ & 0.837$\pm0.058$ & 0.835$\pm0.057$ & 0.837$\pm0.058$ & 0.840$\pm0.041$ \\
          & Logistic Regression & 0.835$\pm0.054$ & 0.835$\pm0.054$ & 0.835$\pm0.054$ & 0.835$\pm0.054$ & 0.907$\pm0.058$ \\
                          & SVM & 0.832$\pm0.036$ & 0.839$\pm0.053$ & 0.840$\pm0.053$ & 0.839$\pm0.053$ & \textbf{0.910$\pm$0.047} \\
       & $\star$ \textbf{Random Forest} & 0.832$\pm$\underline{0.036} & 0.831$\pm$\underline{0.037} & 0.831$\pm$\underline{0.037} & 0.831$\pm$\underline{0.037} & 0.900$\pm$\underline{0.057} \\
                     & AdaBoost & 0.811$\pm0.070$ & 0.807$\pm0.076$ & 0.808$\pm0.075$ & 0.807$\pm0.076$ & 0.881$\pm0.056$ \\
    \midrule
    \multirow{6}{*}{HealthStory} & Classification Layer & 0.529$\pm0.093$ & \textbf{0.717$\pm$0.003} & 0.599$\pm0.008$ & \textbf{0.717$\pm$0.003} & 0.500$\pm0.003$ \\
                         & Naive Bayes & 0.662$\pm0.139$ & 0.600$\pm0.244$ & 0.573$\pm0.289$ & 0.633$\pm0.131$ & 0.508$\pm0.177$ \\
                 & Logistic Regression & 0.660$\pm0.065$ & 0.595$\pm0.206$ & 0.594$\pm0.185$ & 0.584$\pm0.180$ & \textbf{0.557$\pm$0.076} \\
                                 & SVM & 0.649$\pm0.094$ & 0.620$\pm0.137$ & \textbf{0.612$\pm$0.089} & 0.623$\pm0.137$ & 0.536$\pm0.108$ \\
                       & Random Forest & \textbf{0.674$\pm$0.117} & 0.550$\pm0.272$ & 0.526$\pm0.327$ & 0.520$\pm0.269$ & 0.513$\pm0.134$ \\
                            & AdaBoost & 0.656$\pm0.129$ & 0.539$\pm0.302$ & 0.492$\pm0.303$ & 0.540$\pm0.301$ & 0.554$\pm0.076$ \\
    \bottomrule
    \end{tabular}
    \label{t:result_all}
\end{table*}
\subsection{Dataset and Settings}
To test the effectiveness of our method, we conducted our experiments with two real-world datasets: FANG~\cite{nguyen2020fang} and FakeHealth~\cite{dai2020ginger}. 
The dataset FANG was composed in a study by Nguyen et al.~\cite{nguyen2020fang} based on the datasets collected by related work on rumor and news classification~\cite{shu2018fakenewsnet, kochkina_liakata_zubiaga_2018, ma2016detecting}.
% The three main information that are required by \model{} are news urls, corresponding publishers, and the tweet ids about the news. 
The original news content was obtained through the provided news url, and for the 100 news urls that did not have the news content available, resorted to manually searching the news title for the content.
% In the case where the original news was not available, the title itself was used as the news content.
From provided tweet ids, users and their profiles on Twitter could be found through the Twitter API~\cite{twitter_api}.
The labels of the news in FANG are obtained from two well-known fact-checking websites: Snopes~\cite{snopes} and PolitiFact~\cite{politifact}.
FakeHealth is another publicly available benchmark dataset for fake news detection, mainly focused on the healthcare domain.
The dataset consists of two subsets, HealthStory and HealthRelease; HealthStory was used in our study due to the number of news articles in HealthRelease being too small.
HealthStory is collected from the healthcare information review website HealthNewsReviews~\cite{hnr}.
On this website, the professional reviewers gave scores of 1 to 5 for each news.
Similar to the original study that published the FakeHealth dataset, an article is considered as fake if the score is less than three and real otherwise.
The detailed statistics of the dataset used in our experiment are listed in Table~\ref{t:dataset}.
% \footnote{https://github.com/nguyenvanhoang7398/FANG}
% \footnote{https://github.com/EnyanDai/FakeHealth}

\begin{table}[ht]\centering
    \caption{Dataset Statistics.}
    \ra{1.1}
	\begin{tabular}{rrr}
    \toprule
	& \textbf{FANG} & \textbf{HealthStory} \\ 
	\midrule
    \textbf{\# Users} & 52,357 & 63,723 (sampled) \\
    \textbf{\# News} & 1,054 & 1,638 \\
    \textbf{\# of Users per News} & 71.9 & 227.26 \\
    \textbf{\# Fake News} & 448 & 460\\
    \textbf{\# Real News} & 606 & 1,178 \\
    \textbf{\# Publishers} & 442 & 31 \\
    \bottomrule
    \end{tabular}
    \label{t:dataset}
\end{table}

In each dataset, we used 70\% of news articles as our training set, and the remaining 30\% of news articles are further divided into equal sizes of validation and test set. 
For the hyper-parameters, the transformed hidden dimension and the learning rate are set to 512 and 0.0001, respectively. The early-stopping training strategy with patience 20 is adopted to avoid overfitting.
Since fake news detection is a binary classification problem, the real class was treated as positive and the fake class as negative.
% We tested \model{} on Intel (R) Xeon (R) Silver 4214R CPU @ 2.40GHz, 196 GB memory and NVIDIA Titan RTX. 

\subsection{Evaluation of ML Algorithms on News Embedding}

We trained \model{} by connecting the output representation to a fully connected layer to classify the news. 
After training, we evaluated our news representation with five classical machine learning baselines, such as Naive Bayes, Logistic Regression, etc. 
The metrics used for comparison are precision, recall, accuracy, F1 score, and AUC score, and the evaluation results are summarized in Table~\ref{t:result_all}.

As shown in Table~\ref{t:result_all}, the trained classification layer gives relatively better results than other machine learning algorithms in terms of F1 score and accuracy because the classification layer is optimized by classification objective (cross-entropy loss). 
In terms of AUC score, SVM gives a better result, but in terms of standard deviation, random forest generally gives more stable results. 
Based on this, random forest is chosen as the classification algorithm for upcoming evaluations. 
\textbf{Regardless of downstream classification methods, \model{} surpass any existing fake news detection methods} (details in Section~\ref{sec:comp_exist}). 
In the dataset - HealthStory, \model{} does not give an ideal result. 
The explanation for the result on the HealthStory dataset is discussed in the next section.

\subsection{Misinformation vs Disinformation} \label{sec:mis_dis}

Wardle et al. ~\cite{wardle2017information} published a report about information disorder on the Council of Europe in 2017. 
The report intends to examine information disorder and its related challenges.
The authors argue that a large portion of the word 'fake news' consists of three concepts: misinformation, disinformation, and malinformation.
They point out the importance of distinguishing the fake news in accordance with creators' intention and provide the definition of three terms: 

\begin{definition}[\textbf{Disinformation}]
Information that is false and deliberately created to harm a person, social group, organization or country. 
\end{definition}

\begin{definition}[\textbf{Misinformation}]
Information that is false, but not created with the intention of causing harm. 
\end{definition}

\begin{definition}[\textbf{Malinformation}]
Information that is based on reality, used to inflict harm on a person, organization or country. 
\end{definition}

According to the definition of malinformation, it is the information based on reality, while the fake news we talk about in this paper is false information. 
Therefore, we mainly consider disinformation and misinformation here, which are classified according to the news creator's intention.
Considering the definition of fake news given in Section~\ref{sec:preliminaries}, the narrow definition of fake news only covers disinformation, but the broad definition of fake news covers both disinformation and misinformation. 

The dataset FANG is mainly composed of checked news from PolitiFact and Snopes, which are political-related fact-checking websites. 
Thus, the fake news in this dataset is either partisan-biased news or some false information to demean certain politicians, which are considered as information intended to harm the specific person or the organizations. 
Hence, the fake news in this dataset can be considered as \emph{disinformation}. 
The news in HealthStory is collected and fact-checked from Health News Review where evaluates and rates the completeness, accuracy, and balance of news stories that include claims about medical treatments, health care journalism, etc. 
Most of this information is not spread deliberately to harm anyone, so the fake news in the HealthStory dataset can be regarded as \emph{misinformation}.

\begin{figure}[t]
\begin{subfigure}{0.49\columnwidth}
  \centering
  \raisebox{-\height}{\includegraphics[width=\textwidth]{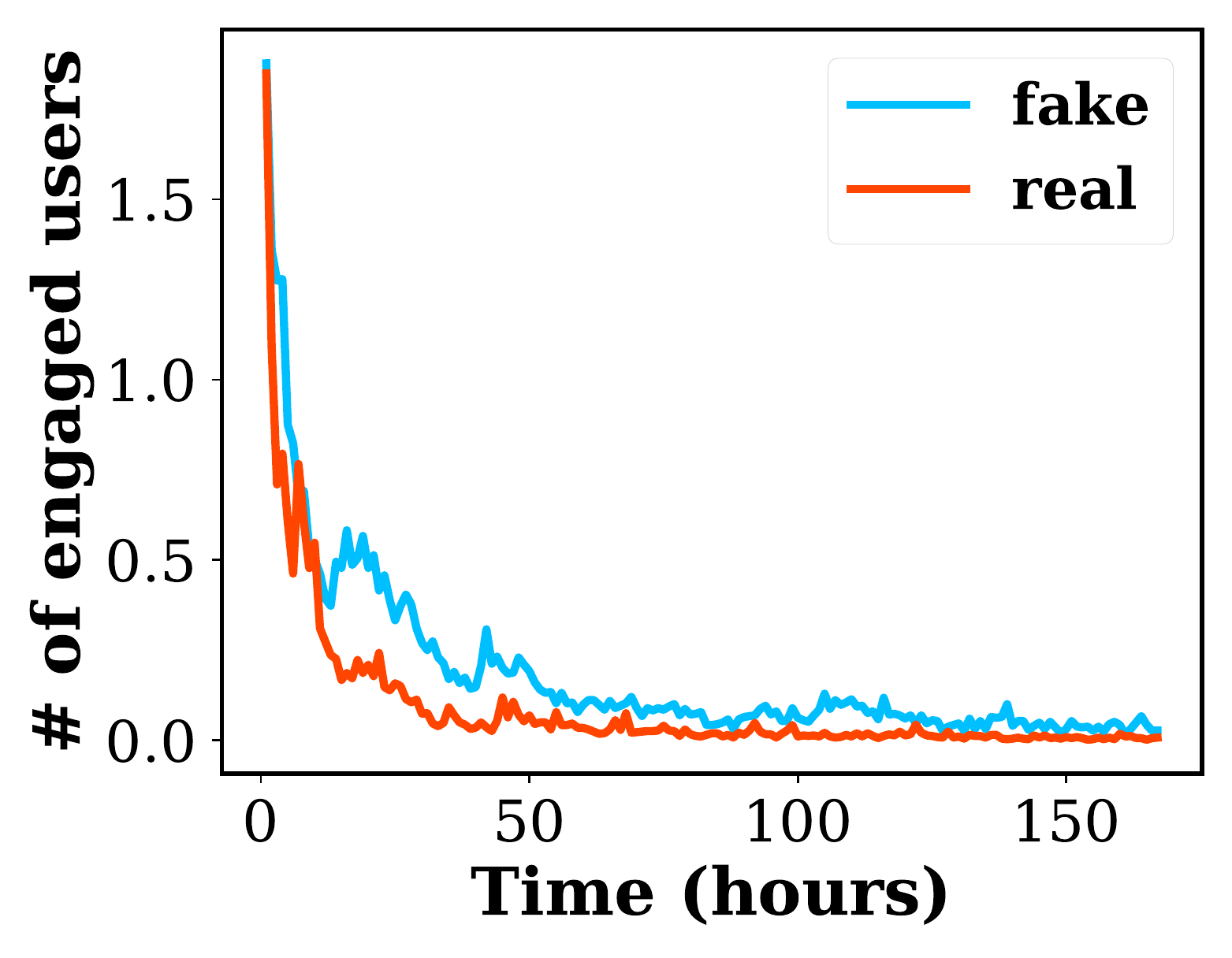}}
%   \caption{}
%   \label{fig:mis_time}
\end{subfigure}
\begin{subfigure}{0.49\columnwidth}
  \centering
  \raisebox{-\height}{\includegraphics[width=\textwidth]{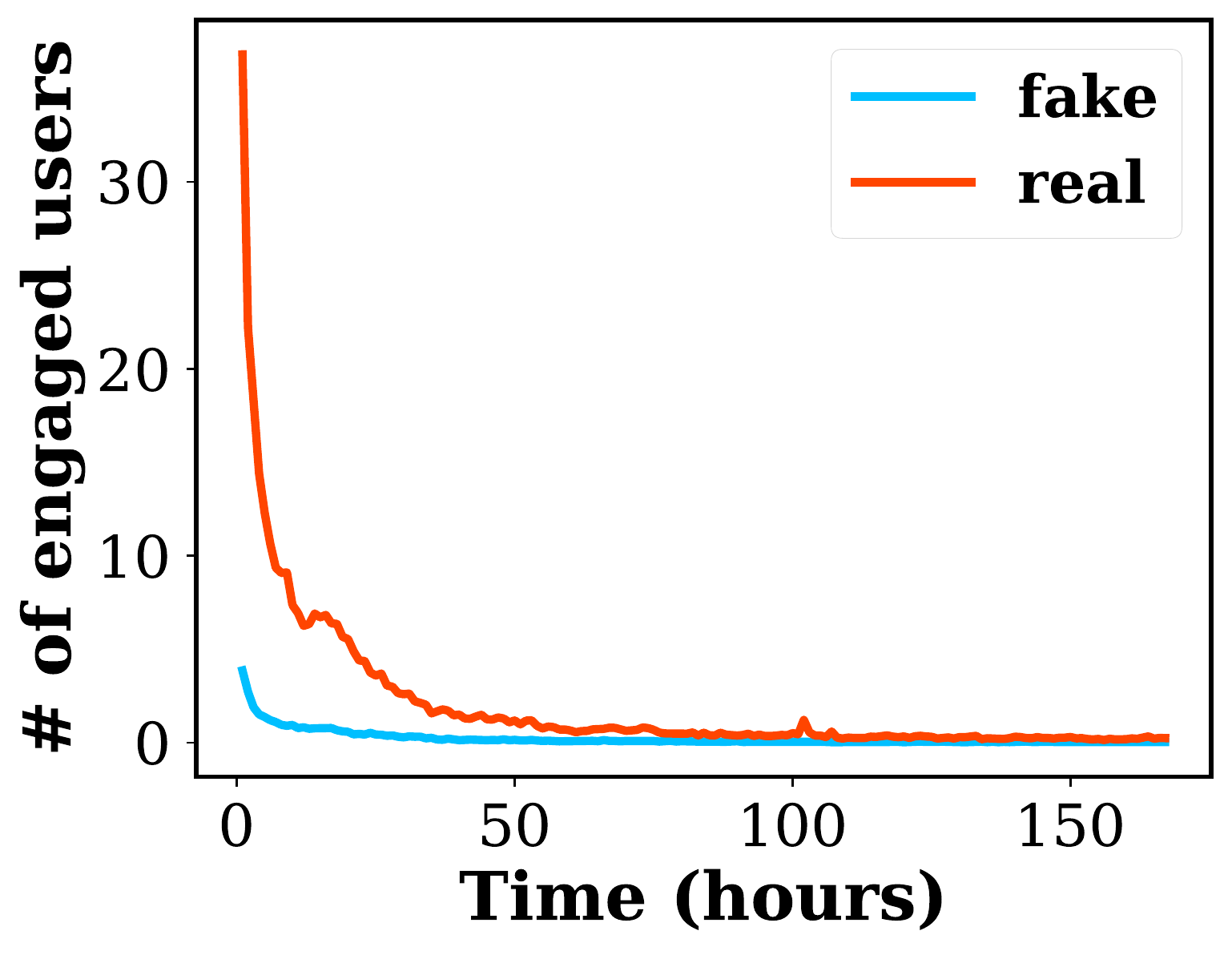}}
%   \caption{}
%   \label{fig:dis_time}
\end{subfigure}
\caption{Comparison of temporal behaviours on two datasets. Both figures show the \# of engagements (tweets) per news vs. time (hours) for FANG (left) and HealthStory (right).}
\label{fig:mis_dis_time}
\vspace{-6pt}
\end{figure}

Figure~\ref{fig:mis_dis_time} compares the number of engaged users along with the time to see how people react to disinformation, misinformation, and real news.
As shown in Figure~\ref{fig:mis_dis_time}, the disinformation (fake in the left) has many periodic spikes, which means the users periodically talk about disinformation.
On the contrary, the misinformation (fake in the right) does not have any periodic spikes and converges to zero not long after the news is published, which is similar to the real news. 
As such, disinformation behaves significantly differently from real information, but misinformation behaves in a similar manner to real news.

To see the impact of temporal information in \model{}, we replace the RNN in \model{} with attention mechanism.
In other words, we checked the detection performance difference between the \model{} with and without temporal information. 
We set the hyperparameters the same for both approaches for a fair comparison, with Random Forest chosen as the classification algorithm. 
The evaluation result on the datasets can be found in Table~\ref{t:mis_dis}. 

% \begin{table}[h]
%     \centering
%     \caption{Performance of the \model{} with and without temporal information.}
%     \ra{1.2}
% 	\begin{tabular}{rrrcrr}
% 	\toprule
% 	& \multicolumn{2}{c}{\textbf{FANG}} & \phantom{abc}& \multicolumn{2}{c}{\textbf{HealthStory}} \\
% 	\cline{2-3} \cline{5-6}
% 	& w/  & w/o  && w/  & w/o  \\
% 	%\midrule
% 	\hline
% 	\textbf{F1 Score} & \textbf{0.831} & 0.759 && 0.526 & \textbf{0.614} \\
% 	\textbf{Accuracy} & \textbf{0.831} & 0.760 && 0.520 & \textbf{0.595} \\
% 	\textbf{AUC score}      & \textbf{0.900} & 0.823 && 0.513 & \textbf{0.636} \\
%     \hline
%     \end{tabular}
%     \label{t:mis_dis}
% \end{table}

\begin{table}[t]
    \centering
    \caption{Performance of the \model{} with and without temporal information.}
    \ra{1.2}
	\begin{tabular}{ccccc}
	\toprule
	\textbf{Dataset} & \textbf{\model{}} & \textbf{F1} & \textbf{Accuracy} & \textbf{AUC} \\
	\midrule
    \multirow{2}{*}{\textbf{FANG}} & \textit{w/ temporal} & \textbf{0.831} & \textbf{0.831} & \textbf{0.900} \\
    & \textit{w/o temporal} & 0.759 & 0.760  & 0.823 \\
	\midrule
    \multirow{2}{*}{\textbf{HealthStory}} & \textit{w/ temporal} & 0.526 & 0.520 & 0.513  \\
    & \textit{w/o temporal} & \textbf{0.614} & \textbf{0.595} & \textbf{0.636} \\
    \bottomrule
    \end{tabular}
    \label{t:mis_dis}
    \vspace{-6pt}
\end{table}

The results show that the RNN based approach performs better than the other one in FANG dataset, but for the HealthStory dataset, the performance is better when the attention is applied. 
This means the existence of temporal information is not helpful in detecting misinformation. 
Furthermore, in FANG dataset, the validation loss of \model{} with RNN converges much faster than the one with attention mechanism; by contrast, the convergence speed of the two approaches is similar in the HealthStory dataset. (See Figure~\ref{fig:mis_dis_val} in Appendix)

To sum up, in a dataset has temporal behavior difference between real and fake class (i.e., disinformation dataset), \model{} with RNN not only improves the performance of the fake news detection but also accelerates the learning speed.
% News in HealthStory has no clear distinction between real and fake in terms of the temporal behavior of user engagement.
% \jian{Thus, \model{} that aims to capture social context and temporal information of social user engagement is hard to identify misinformation.
% Therefore, we mainly focus on the disinformation dataset, FANG, in the remaining part of the experiments.}

\subsection{Comparison with Existing Methods} \label{sec:comp_exist}

To show that \model{} is superior to other fake news detection, we compared \model{} with other existing fake news detection methods.
The bench-marked detection methods can be categorized into text-based approaches and graph-based approaches.
For text-based approach, we use three different document embedding methods, TF-IDF, LIWC~\cite{pennebaker2015development}, and Doc2Vec~\cite{le2014distributed}, combined with SVM as baselines; and several representative graph-based fake news detection frameworks~\cite{chandra2020graph, ruchansky2017csi, nguyen2020fang, ren2020adversarial} are also compared in this experiment.

\model{} is also compared with some Graph Neural Network (GNN) methods to show that \model{} is better than just simply applying the GNN on the graph.
The basic GNN methods~\cite{kipf2016semi, velivckovic2017graph, hamilton2017inductive}, as well as the methods tailor to the heterogeneous graph, are compared ~\cite{han2020graph, schlichtkrull2018modeling}.
The brief descriptions of the aforementioned fake news detection methods and GNN baselines we compared with are listed in the Appendix.

\begin{table}[t]\centering
    \caption{Comparison with other methods. The AUC score of the CSI is from FANG, the F1 score and AUC score are not reported in this paper.}
    \ra{1.2}
	\begin{tabular}{crccc}
	\toprule
	 \textbf{Category} & \textbf{Method} & \textbf{F1} & \textbf{Accuracy} & \textbf{AUC} \\ 
	\midrule    
	\multirowcell{3}{Text-\\based} & 
	TF.IDF + SVM & 0.746 & 0.750 & 0.735 \\
    & LIWC + SVM & 0.512 & 0.550 &  0.511\\
    & Doc2Vec + SVM & 0.561 & 0.560 & 0.554\\
    \cline{1-5}
    \multirowcell{4}{Graph-\\based} & 
    CSI & - & - &  0.741 \\
    & SAFER & 0.678 & 0.680 & 0.669 \\
    & FANG & 0.676 & 0.687 & 0.750 \\
    & AA-HGNN & 0.726 & 0.662 & 0.654 \\
    \cline{1-5}
    \multirowcell{6}{GNN-\\baselines} &
    GCN        & 0.645 & 0.650 & 0.633 \\
    &GAT        & 0.642 & 0.650 & 0.630 \\
    &GraphSAGE  & 0.779 & 0.780 & 0.773 \\
    % \cline{2-5}
    &R-GCN      & 0.765 & 0.770 & 0.753 \\
    &HAN        & 0.662 & 0.660 & 0.658 \\
    % &\jian{MAGNN} & \jian{0.761} & \jian{0.787} & \jian{0.843} \\
    \cline{1-5}
	\multicolumn{2}{c}{\textbf{\model{}}} & \textbf{0.831} & \textbf{0.831} & \textbf{0.900} \\
    \bottomrule
    \end{tabular}
    \label{t:vs_other_methods}
    \vspace{-10pt}
\end{table}

The results of  Table~\ref{t:vs_other_methods} indicates that \model{} outperforms existing text-based or graph-based fake news detection methods. 
This is because these existing approaches cannot produce representation with rich social context and temporal information as \model{} do, i.e., they fail to tackle all three difficulties.
CSI and SAFER, for example, did not use multi-level social context, and they also incurred some information loss as they ignored the node and relation types.
AA-HGNN, including SAFER, miss temporal information in the news representation.
AA-HGNN also did not use users as social context.
FANG performs better than these methods since it tries to preserve multi-level social context and temporal information.
To preserve information, FANG divides the fake news detection task into several sub-tasks, and each sub-task deals with certain information.
Dividing into several sub-tasks is ineffective because errors in sub-task will be propagated up to the final news representation and thus harm the detection performance.
As such, the result emphasizes the importance of resolving the proposed three difficulties in fake news detection.

For GNN baselines, the graph embedding methods made for homogeneous graphs, such as GCN, GAT, and GraphSAGE, did not give ideal results since node types and relations are ignored in these cases.
R-GCN and HAN, which are designed for heterogeneous graph, also has no significant improvement, which implies that \model{} is better than a simple application of these graph embedding methods on the heterogeneous graph of news.
The fail of GNN baselines target on the heterogeneous graph can attribute to the missing temporal information of user engagement, which is the second difficulty that needs to be resolved in the social context-aware fake news detection.

\subsection{Limited training data}

Normally, the fake news dataset has limited training data due to the large-scale requirement of human labor, so the model should work well in the circumstance of limited training samples. To show that \model{} outperforms existing methods given the circumstance of scarce training data, we gradually enlarge the training data, from 10\% to 90\%, and compare the fake news detection result with existing methods. Table~\ref{t:train_ratio} shows the comparison result.  

\begin{table}[h]
    \center
    \caption{Comparison of AUC score against other fake news detection methods by varying the size of the training data.}
	\begin{tabular}{rccccc}
	\toprule
	& \textbf{10\%} & \textbf{30\%} & \textbf{50\%} & \textbf{70\%} & \textbf{90\%}\\ 
	\midrule
    CSI   & 0.636 & 0.671 & 0.670 & 0.689 & 0.691 \\
    SAFER  & 0.546 & 0.689 & 0.666 & 0.692 & 0.669 \\
    FANG & 0.669 & 0.704 & 0.717 & 0.723 & 0.752 \\
    AA-HGNN & 0.573  & 0.598 & 0.656 & 0.657 & 0.642 \\
    $\text{\model{}}_{w/o\ time}$ & 0.594 & 0.707 & 0.776 & 0.749 & 0.751\\
    \textbf{$\text{\model{}}_{w/\ time}$} & \textbf{0.764} & \textbf{0.835} & \textbf{0.878} & \textbf{0.889} & \textbf{0.900} \\
    \bottomrule
    \end{tabular}
    \label{t:train_ratio}
    \vspace{-13pt}
\end{table}

The AUC score of \model{} achieves over 0.8 with only 30\% of training data and even outperforms the rest of the methods with 90\% of the training data. 
AA-HGNN is designed to overcome the scarcity of training data issues in the fake news detection task, but \model{} is still better than AA-HGNN even when the size of training data is small.

% \begin{figure}[ht]
%     \centering
%     \includegraphics[width=\columnwidth]{figure/train_ratio.pdf} 
%     \caption{Comparison of AUC score against other fake news detection methods by varying the size of the training data.}
%     \label{fig:train_ratio}
% \end{figure}

\section{Discussion}

\subsection{Inductiveness of \model{}}
A deep learning based approach dealing with graph-structured data should have generality to produce practical predictions for unseen data.
A method is an inductive approach if it can generate embeddings for the nodes that were not seen during training. 
In contrast, it is called a transductive approach if the method cannot generate embeddings for the nodes appearing in the testing phase for the first time. 
For example, GCN is inductive, whereas Node2Vec is transductive.

In graph-based fake news detection, unseen nodes can appear in the testing phase. 
It might be newly published news, new publishers, or new users.
Some approaches using matrix decomposition~\cite{ruchansky2017csi, shu2019beyond} are not able to generate embedding for newly published news with social context information.
In \model{}, however, the learnable parameters in our model are used after Meta-Path extraction with random sampling, and they are shared by all nodes. 
Therefore, our method is highly inductive, that is, \model{} can generate news embeddings that are not seen during the training. 

\subsection{Limitation and Future Work}
As expected, a single news article is engaged with by a large number of users.
Using every single user's information as a feature is therefore impractical, and we eventually used simple random sampling to select a certain number of users.
Therefore, an improved method of screening important users is necessary for fake news detection to overcome the limitation.
In addition, to apply the proposed method, we must first identify the relevant tweets for particular news.
Since this paper focuses primarily on identifying the news in the context in which news and related tweets are given, finding relevant tweets for particular news is left as future work.
\section{Conclusions}
Fake news is a critical social problem threatening many aspects of the lives of the general public. 
We pose three difficulties in social context aware fake news detection and address them by proposing a novel fake news detection framework ~\model{}.
Our model overcomes the shortcomings of the previous graph-based approaches and exhibits state-of-the-art performance.
We also provide insight about misinformation and disinformation by clarifying their different propagation properties.
\model{} can be of aid in future studies not only residing to fake news detection but also various events concerning disinformation.

%Note that in the new ACM style, the Appendices come before the References.

\begin{acks}
% TODO: For the submission, don't include acknowledgments since they would most likely deanonymize you.
\end{acks}

\bibliographystyle{ACM-Reference-Format}
\bibliography{main}


\begin{thebibliography}{52}

%%% the Web address.

\ifx \showCODEN    \undefined \def \showCODEN     #1{\unskip}     \fi
\ifx \showDOI      \undefined \def \showDOI       #1{#1}\fi
\ifx \showISBNx    \undefined \def \showISBNx     #1{\unskip}     \fi
\ifx \showISBNxiii \undefined \def \showISBNxiii  #1{\unskip}     \fi
\ifx \showISSN     \undefined \def \showISSN      #1{\unskip}     \fi
\ifx \showLCCN     \undefined \def \showLCCN      #1{\unskip}     \fi
\ifx \shownote     \undefined \def \shownote      #1{#1}          \fi
\ifx \showarticletitle \undefined \def \showarticletitle #1{#1}   \fi
\ifx \showURL      \undefined \def \showURL       {\relax}        \fi
% invisible to TeX
\providecommand\bibfield[2]{#2}
\providecommand\bibinfo[2]{#2}
\providecommand\natexlab[1]{#1}
\providecommand\showeprint[2][]{arXiv:#2}

\bibitem[\protect\citeauthoryear{??}{cov}{2020}]%
        {covid_rumours}
 \bibinfo{year}{2020}\natexlab{}.
\newblock \bibinfo{title}{Coronavirus: The viral rumours that were completely
  wrong}.
\newblock
  \bibinfo{howpublished}{\url{https://www.bbc.com/news/blogs-trending-53640964}}.
\newblock


\bibitem[\protect\citeauthoryear{??}{hnr}{2020}]%
        {hnr}
 \bibinfo{year}{2020}\natexlab{}.
\newblock \bibinfo{title}{Health News Review}.
\newblock \bibinfo{howpublished}{\url{https://www.healthnewsreview.org/}}.
\newblock


\bibitem[\protect\citeauthoryear{??}{mbf}{2020a}]%
        {mbfc_left}
 \bibinfo{year}{2020}\natexlab{a}.
\newblock \bibinfo{title}{Left bias Publishers checked by MBFC}.
\newblock \bibinfo{howpublished}{\url{https://mediabiasfactcheck.com/left/}}.
\newblock


\bibitem[\protect\citeauthoryear{??}{pol}{2020}]%
        {politifact}
 \bibinfo{year}{2020}\natexlab{}.
\newblock \bibinfo{title}{PolitiFact}.
\newblock \bibinfo{howpublished}{\url{https://www.politifact.com/}}.
\newblock


\bibitem[\protect\citeauthoryear{??}{mbf}{2020b}]%
        {mbfc_source}
 \bibinfo{year}{2020}\natexlab{b}.
\newblock \bibinfo{title}{Questionable Publishers checked by MBFC}.
\newblock
  \bibinfo{howpublished}{\url{https://mediabiasfactcheck.com/fake-news/}}.
\newblock


\bibitem[\protect\citeauthoryear{??}{mbf}{2020c}]%
        {mbfc_right}
 \bibinfo{year}{2020}\natexlab{c}.
\newblock \bibinfo{title}{Right bias Publishers checked by MBFC}.
\newblock \bibinfo{howpublished}{\url{https://mediabiasfactcheck.com/right/}}.
\newblock


\bibitem[\protect\citeauthoryear{??}{sno}{2020}]%
        {snopes}
 \bibinfo{year}{2020}\natexlab{}.
\newblock \bibinfo{title}{Snopes}.
\newblock \bibinfo{howpublished}{\url{https://www.snopes.com/}}.
\newblock


\bibitem[\protect\citeauthoryear{??}{twi}{2020}]%
        {twitter_api}
 \bibinfo{year}{2020}\natexlab{}.
\newblock \bibinfo{title}{Twitter API}.
\newblock
  \bibinfo{howpublished}{\url{https://developer.twitter.com/en/docs/twitter-api}}.
\newblock


\bibitem[\protect\citeauthoryear{??}{ele}{2020}]%
        {election_fraud}
 \bibinfo{year}{2020}\natexlab{}.
\newblock \bibinfo{title}{US election 2020: Fact-checking Trump team's main
  fraud claims}.
\newblock
  \bibinfo{howpublished}{\url{https://www.bbc.com/news/election-us-2020-55016029}}.
\newblock


\bibitem[\protect\citeauthoryear{??}{fb_}{2021}]%
        {fb_report}
 \bibinfo{year}{2021}\natexlab{}.
\newblock \bibinfo{title}{Facebook Media: Working to Stop Misinformation and
  False News}.
\newblock
  \bibinfo{howpublished}{\url{https://www.facebook.com/formedia/blog/working-to-stop-misinformation-and-false-news}}.
\newblock


\bibitem[\protect\citeauthoryear{Bian, Xiao, Xu, Zhao, Huang, Rong, and
  Huang}{Bian et~al\mbox{.}}{2020}]%
        {bian2020rumor}
\bibfield{author}{\bibinfo{person}{Tian Bian}, \bibinfo{person}{Xi Xiao},
  \bibinfo{person}{Tingyang Xu}, \bibinfo{person}{Peilin Zhao},
  \bibinfo{person}{Wenbing Huang}, \bibinfo{person}{Yu Rong}, {and}
  \bibinfo{person}{Junzhou Huang}.} \bibinfo{year}{2020}\natexlab{}.
\newblock \showarticletitle{Rumor detection on social media with bi-directional
  graph convolutional networks}. In \bibinfo{booktitle}{\emph{Proceedings of
  the AAAI Conference on Artificial Intelligence}}, Vol.~\bibinfo{volume}{34}.
  \bibinfo{pages}{549--556}.
\newblock


\bibitem[\protect\citeauthoryear{Castillo, Mendoza, and Poblete}{Castillo
  et~al\mbox{.}}{2011}]%
        {castillo2011information}
\bibfield{author}{\bibinfo{person}{Carlos Castillo}, \bibinfo{person}{Marcelo
  Mendoza}, {and} \bibinfo{person}{Barbara Poblete}.}
  \bibinfo{year}{2011}\natexlab{}.
\newblock \showarticletitle{Information credibility on twitter}. In
  \bibinfo{booktitle}{\emph{Proceedings of the 20th international conference on
  World wide web}}. \bibinfo{pages}{675--684}.
\newblock


\bibitem[\protect\citeauthoryear{Chandra, Mishra, Yannakoudakis, and
  Shutova}{Chandra et~al\mbox{.}}{2020}]%
        {chandra2020graph}
\bibfield{author}{\bibinfo{person}{Shantanu Chandra}, \bibinfo{person}{Pushkar
  Mishra}, \bibinfo{person}{Helen Yannakoudakis}, {and}
  \bibinfo{person}{Ekaterina Shutova}.} \bibinfo{year}{2020}\natexlab{}.
\newblock \showarticletitle{Graph-based Modeling of Online Communities for Fake
  News Detection}.
\newblock \bibinfo{journal}{\emph{arXiv preprint arXiv:2008.06274}}
  (\bibinfo{year}{2020}).
\newblock


\bibitem[\protect\citeauthoryear{Chung, Gulcehre, Cho, and Bengio}{Chung
  et~al\mbox{.}}{2014}]%
        {chung2014empirical}
\bibfield{author}{\bibinfo{person}{Junyoung Chung}, \bibinfo{person}{Caglar
  Gulcehre}, \bibinfo{person}{KyungHyun Cho}, {and} \bibinfo{person}{Yoshua
  Bengio}.} \bibinfo{year}{2014}\natexlab{}.
\newblock \showarticletitle{Empirical evaluation of gated recurrent neural
  networks on sequence modeling}.
\newblock \bibinfo{journal}{\emph{arXiv preprint arXiv:1412.3555}}
  (\bibinfo{year}{2014}).
\newblock


\bibitem[\protect\citeauthoryear{Conroy, Rubin, and Chen}{Conroy
  et~al\mbox{.}}{2015}]%
        {conroy2015automatic}
\bibfield{author}{\bibinfo{person}{Niall~J Conroy}, \bibinfo{person}{Victoria~L
  Rubin}, {and} \bibinfo{person}{Yimin Chen}.} \bibinfo{year}{2015}\natexlab{}.
\newblock \showarticletitle{Automatic deception detection: methods for finding
  fake news}. In \bibinfo{booktitle}{\emph{Proceedings of the 78th ASIS\&T
  Annual Meeting: Information Science with Impact: Research in and for the
  Community}}. \bibinfo{pages}{1--4}.
\newblock


\bibitem[\protect\citeauthoryear{Dai, Sun, and Wang}{Dai et~al\mbox{.}}{2020}]%
        {dai2020ginger}
\bibfield{author}{\bibinfo{person}{Enyan Dai}, \bibinfo{person}{Yiwei Sun},
  {and} \bibinfo{person}{Suhang Wang}.} \bibinfo{year}{2020}\natexlab{}.
\newblock \showarticletitle{Ginger cannot cure cancer: Battling fake health
  news with a comprehensive data repository}. In
  \bibinfo{booktitle}{\emph{Proceedings of the International AAAI Conference on
  Web and Social Media}}, Vol.~\bibinfo{volume}{14}. \bibinfo{pages}{853--862}.
\newblock


\bibitem[\protect\citeauthoryear{Dettmers, Minervini, Stenetorp, and
  Riedel}{Dettmers et~al\mbox{.}}{2018}]%
        {dettmers2018convolutional}
\bibfield{author}{\bibinfo{person}{Tim Dettmers}, \bibinfo{person}{Pasquale
  Minervini}, \bibinfo{person}{Pontus Stenetorp}, {and}
  \bibinfo{person}{Sebastian Riedel}.} \bibinfo{year}{2018}\natexlab{}.
\newblock \showarticletitle{Convolutional 2d knowledge graph embeddings}. In
  \bibinfo{booktitle}{\emph{Proceedings of the AAAI Conference on Artificial
  Intelligence}}, Vol.~\bibinfo{volume}{32}.
\newblock


\bibitem[\protect\citeauthoryear{Feng, Banerjee, and Choi}{Feng
  et~al\mbox{.}}{2012}]%
        {feng2012syntactic}
\bibfield{author}{\bibinfo{person}{Song Feng}, \bibinfo{person}{Ritwik
  Banerjee}, {and} \bibinfo{person}{Yejin Choi}.}
  \bibinfo{year}{2012}\natexlab{}.
\newblock \showarticletitle{Syntactic stylometry for deception detection}. In
  \bibinfo{booktitle}{\emph{Proceedings of the 50th Annual Meeting of the
  Association for Computational Linguistics (Volume 2: Short Papers)}}.
  \bibinfo{pages}{171--175}.
\newblock


\bibitem[\protect\citeauthoryear{Fu, Zhang, Meng, and King}{Fu
  et~al\mbox{.}}{2020}]%
        {fu2020magnn}
\bibfield{author}{\bibinfo{person}{Xinyu Fu}, \bibinfo{person}{Jiani Zhang},
  \bibinfo{person}{Ziqiao Meng}, {and} \bibinfo{person}{Irwin King}.}
  \bibinfo{year}{2020}\natexlab{}.
\newblock \showarticletitle{MAGNN: metapath aggregated graph neural network for
  heterogeneous graph embedding}. In \bibinfo{booktitle}{\emph{Proceedings of
  The Web Conference 2020}}. \bibinfo{pages}{2331--2341}.
\newblock


\bibitem[\protect\citeauthoryear{Giachanou, Zhang, and Rosso}{Giachanou
  et~al\mbox{.}}{2020}]%
        {giachanou2020multimodal}
\bibfield{author}{\bibinfo{person}{Anastasia Giachanou},
  \bibinfo{person}{Guobiao Zhang}, {and} \bibinfo{person}{Paolo Rosso}.}
  \bibinfo{year}{2020}\natexlab{}.
\newblock \showarticletitle{Multimodal Fake News Detection with Textual, Visual
  and Semantic Information}. In \bibinfo{booktitle}{\emph{International
  Conference on Text, Speech, and Dialogue}}. Springer,
  \bibinfo{pages}{30--38}.
\newblock


\bibitem[\protect\citeauthoryear{Grover and Leskovec}{Grover and
  Leskovec}{2016}]%
        {grover2016node2vec}
\bibfield{author}{\bibinfo{person}{Aditya Grover} {and} \bibinfo{person}{Jure
  Leskovec}.} \bibinfo{year}{2016}\natexlab{}.
\newblock \showarticletitle{node2vec: Scalable feature learning for networks}.
  In \bibinfo{booktitle}{\emph{Proceedings of the 22nd ACM SIGKDD international
  conference on Knowledge discovery and data mining}}.
  \bibinfo{pages}{855--864}.
\newblock


\bibitem[\protect\citeauthoryear{Hamilton, Ying, and Leskovec}{Hamilton
  et~al\mbox{.}}{2017}]%
        {hamilton2017inductive}
\bibfield{author}{\bibinfo{person}{William~L. Hamilton}, \bibinfo{person}{Rex
  Ying}, {and} \bibinfo{person}{Jure Leskovec}.}
  \bibinfo{year}{2017}\natexlab{}.
\newblock \showarticletitle{Inductive Representation Learning on Large Graphs}.
  In \bibinfo{booktitle}{\emph{NIPS}}.
\newblock


\bibitem[\protect\citeauthoryear{Han, Karunasekera, and Leckie}{Han
  et~al\mbox{.}}{2020}]%
        {han2020graph}
\bibfield{author}{\bibinfo{person}{Yi Han}, \bibinfo{person}{Shanika
  Karunasekera}, {and} \bibinfo{person}{Christopher Leckie}.}
  \bibinfo{year}{2020}\natexlab{}.
\newblock \showarticletitle{Graph neural networks with continual learning for
  fake news detection from social media}.
\newblock \bibinfo{journal}{\emph{arXiv preprint arXiv:2007.03316}}
  (\bibinfo{year}{2020}).
\newblock


\bibitem[\protect\citeauthoryear{Khattar, Goud, Gupta, and Varma}{Khattar
  et~al\mbox{.}}{2019}]%
        {khattar2019mvae}
\bibfield{author}{\bibinfo{person}{Dhruv Khattar},
  \bibinfo{person}{Jaipal~Singh Goud}, \bibinfo{person}{Manish Gupta}, {and}
  \bibinfo{person}{Vasudeva Varma}.} \bibinfo{year}{2019}\natexlab{}.
\newblock \showarticletitle{Mvae: Multimodal variational autoencoder for fake
  news detection}. In \bibinfo{booktitle}{\emph{The World Wide Web
  Conference}}. \bibinfo{pages}{2915--2921}.
\newblock


\bibitem[\protect\citeauthoryear{Kipf and Welling}{Kipf and Welling}{2016}]%
        {kipf2016semi}
\bibfield{author}{\bibinfo{person}{Thomas~N Kipf} {and} \bibinfo{person}{Max
  Welling}.} \bibinfo{year}{2016}\natexlab{}.
\newblock \showarticletitle{Semi-supervised classification with graph
  convolutional networks}.
\newblock \bibinfo{journal}{\emph{arXiv preprint arXiv:1609.02907}}
  (\bibinfo{year}{2016}).
\newblock


\bibitem[\protect\citeauthoryear{Kochkina, Liakata, and Zubiaga}{Kochkina
  et~al\mbox{.}}{2018}]%
        {kochkina_liakata_zubiaga_2018}
\bibfield{author}{\bibinfo{person}{Elena Kochkina}, \bibinfo{person}{Maria
  Liakata}, {and} \bibinfo{person}{Arkaitz Zubiaga}.}
  \bibinfo{year}{2018}\natexlab{}.
\newblock \bibinfo{title}{PHEME dataset for Rumour Detection and Veracity
  Classification}.
\newblock
\newblock
\urldef\tempurl%
\url{https://doi.org/10.6084/m9.figshare.6392078.v1}
\showDOI{\tempurl}


\bibitem[\protect\citeauthoryear{Kwon, Cha, Jung, Chen, and Wang}{Kwon
  et~al\mbox{.}}{2013}]%
        {kwon2013prominent}
\bibfield{author}{\bibinfo{person}{Sejeong Kwon}, \bibinfo{person}{Meeyoung
  Cha}, \bibinfo{person}{Kyomin Jung}, \bibinfo{person}{Wei Chen}, {and}
  \bibinfo{person}{Yajun Wang}.} \bibinfo{year}{2013}\natexlab{}.
\newblock \showarticletitle{Prominent features of rumor propagation in online
  social media}. In \bibinfo{booktitle}{\emph{2013 IEEE 13th international
  conference on data mining}}. IEEE, \bibinfo{pages}{1103--1108}.
\newblock


\bibitem[\protect\citeauthoryear{Le and Mikolov}{Le and Mikolov}{2014}]%
        {le2014distributed}
\bibfield{author}{\bibinfo{person}{Quoc Le} {and} \bibinfo{person}{Tomas
  Mikolov}.} \bibinfo{year}{2014}\natexlab{}.
\newblock \showarticletitle{Distributed representations of sentences and
  documents}. In \bibinfo{booktitle}{\emph{International conference on machine
  learning}}. PMLR, \bibinfo{pages}{1188--1196}.
\newblock


\bibitem[\protect\citeauthoryear{Ma, Gao, Mitra, Kwon, Jansen, Wong, and
  Cha}{Ma et~al\mbox{.}}{2016}]%
        {ma2016detecting}
\bibfield{author}{\bibinfo{person}{Jing Ma}, \bibinfo{person}{Wei Gao},
  \bibinfo{person}{Prasenjit Mitra}, \bibinfo{person}{Sejeong Kwon},
  \bibinfo{person}{Bernard~J Jansen}, \bibinfo{person}{Kam-Fai Wong}, {and}
  \bibinfo{person}{Meeyoung Cha}.} \bibinfo{year}{2016}\natexlab{}.
\newblock \showarticletitle{Detecting rumors from microblogs with recurrent
  neural networks}.
\newblock  (\bibinfo{year}{2016}).
\newblock


\bibitem[\protect\citeauthoryear{Maaten and Hinton}{Maaten and Hinton}{2008}]%
        {maaten2008visualizing}
\bibfield{author}{\bibinfo{person}{Laurens van~der Maaten} {and}
  \bibinfo{person}{Geoffrey Hinton}.} \bibinfo{year}{2008}\natexlab{}.
\newblock \showarticletitle{Visualizing data using t-SNE}.
\newblock \bibinfo{journal}{\emph{Journal of machine learning research}}
  \bibinfo{volume}{9}, \bibinfo{number}{Nov} (\bibinfo{year}{2008}),
  \bibinfo{pages}{2579--2605}.
\newblock


\bibitem[\protect\citeauthoryear{Mikolov, Chen, Corrado, and Dean}{Mikolov
  et~al\mbox{.}}{2013}]%
        {mikolov2013efficient}
\bibfield{author}{\bibinfo{person}{Tomas Mikolov}, \bibinfo{person}{Kai Chen},
  \bibinfo{person}{Greg Corrado}, {and} \bibinfo{person}{Jeffrey Dean}.}
  \bibinfo{year}{2013}\natexlab{}.
\newblock \showarticletitle{Efficient estimation of word representations in
  vector space}.
\newblock \bibinfo{journal}{\emph{arXiv preprint arXiv:1301.3781}}
  (\bibinfo{year}{2013}).
\newblock


\bibitem[\protect\citeauthoryear{Nguyen, Sugiyama, Nakov, and Kan}{Nguyen
  et~al\mbox{.}}{2020}]%
        {nguyen2020fang}
\bibfield{author}{\bibinfo{person}{Van-Hoang Nguyen}, \bibinfo{person}{Kazunari
  Sugiyama}, \bibinfo{person}{Preslav Nakov}, {and} \bibinfo{person}{Min-Yen
  Kan}.} \bibinfo{year}{2020}\natexlab{}.
\newblock \showarticletitle{FANG: Leveraging social context for fake news
  detection using graph representation}. In
  \bibinfo{booktitle}{\emph{Proceedings of the 29th ACM International
  Conference on Information \& Knowledge Management}}.
  \bibinfo{pages}{1165--1174}.
\newblock


\bibitem[\protect\citeauthoryear{Pennebaker, Boyd, Jordan, and
  Blackburn}{Pennebaker et~al\mbox{.}}{2015}]%
        {pennebaker2015development}
\bibfield{author}{\bibinfo{person}{James~W Pennebaker}, \bibinfo{person}{Ryan~L
  Boyd}, \bibinfo{person}{Kayla Jordan}, {and} \bibinfo{person}{Kate
  Blackburn}.} \bibinfo{year}{2015}\natexlab{}.
\newblock \bibinfo{booktitle}{\emph{The development and psychometric properties
  of LIWC2015}}.
\newblock \bibinfo{type}{{T}echnical {R}eport}.
\newblock


\bibitem[\protect\citeauthoryear{P{\'e}rez-Rosas, Kleinberg, Lefevre, and
  Mihalcea}{P{\'e}rez-Rosas et~al\mbox{.}}{2018}]%
        {perez2018automatic}
\bibfield{author}{\bibinfo{person}{Ver{\'o}nica P{\'e}rez-Rosas},
  \bibinfo{person}{Bennett Kleinberg}, \bibinfo{person}{Alexandra Lefevre},
  {and} \bibinfo{person}{Rada Mihalcea}.} \bibinfo{year}{2018}\natexlab{}.
\newblock \showarticletitle{Automatic Detection of Fake News}. In
  \bibinfo{booktitle}{\emph{Proceedings of the 27th International Conference on
  Computational Linguistics}}. \bibinfo{pages}{3391--3401}.
\newblock


\bibitem[\protect\citeauthoryear{Potthast, Kiesel, Reinartz, Bevendorff, and
  Stein}{Potthast et~al\mbox{.}}{2018}]%
        {potthast2018stylometric}
\bibfield{author}{\bibinfo{person}{Martin Potthast}, \bibinfo{person}{Johannes
  Kiesel}, \bibinfo{person}{Kevin Reinartz}, \bibinfo{person}{Janek
  Bevendorff}, {and} \bibinfo{person}{Benno Stein}.}
  \bibinfo{year}{2018}\natexlab{}.
\newblock \showarticletitle{A Stylometric Inquiry into Hyperpartisan and Fake
  News}. In \bibinfo{booktitle}{\emph{Proceedings of the 56th Annual Meeting of
  the Association for Computational Linguistics (Volume 1: Long Papers)}}.
  \bibinfo{pages}{231--240}.
\newblock


\bibitem[\protect\citeauthoryear{Qian, Wang, Hu, Fang, and Xu}{Qian
  et~al\mbox{.}}{2021}]%
        {qian2021hierarchical}
\bibfield{author}{\bibinfo{person}{Shengsheng Qian}, \bibinfo{person}{Jinguang
  Wang}, \bibinfo{person}{Jun Hu}, \bibinfo{person}{Quan Fang}, {and}
  \bibinfo{person}{Changsheng Xu}.} \bibinfo{year}{2021}\natexlab{}.
\newblock \showarticletitle{Hierarchical multi-modal contextual attention
  network for fake news detection}. In \bibinfo{booktitle}{\emph{Proceedings of
  the 44th International ACM SIGIR Conference on Research and Development in
  Information Retrieval}}. \bibinfo{pages}{153--162}.
\newblock


\bibitem[\protect\citeauthoryear{Ren, Wang, Zhang, and Chang}{Ren
  et~al\mbox{.}}{2020}]%
        {ren2020adversarial}
\bibfield{author}{\bibinfo{person}{Yuxiang Ren}, \bibinfo{person}{Bo Wang},
  \bibinfo{person}{Jiawei Zhang}, {and} \bibinfo{person}{Yi Chang}.}
  \bibinfo{year}{2020}\natexlab{}.
\newblock \showarticletitle{Adversarial active learning based heterogeneous
  graph neural network for fake news detection}. In
  \bibinfo{booktitle}{\emph{2020 IEEE International Conference on Data Mining
  (ICDM)}}. IEEE, \bibinfo{pages}{452--461}.
\newblock


\bibitem[\protect\citeauthoryear{Rubin and Lukoianova}{Rubin and
  Lukoianova}{2015}]%
        {rubin2015truth}
\bibfield{author}{\bibinfo{person}{Victoria~L Rubin} {and}
  \bibinfo{person}{Tatiana Lukoianova}.} \bibinfo{year}{2015}\natexlab{}.
\newblock \showarticletitle{Truth and deception at the rhetorical structure
  level}.
\newblock \bibinfo{journal}{\emph{Journal of the Association for Information
  Science and Technology}} \bibinfo{volume}{66}, \bibinfo{number}{5}
  (\bibinfo{year}{2015}), \bibinfo{pages}{905--917}.
\newblock


\bibitem[\protect\citeauthoryear{Ruchansky, Seo, and Liu}{Ruchansky
  et~al\mbox{.}}{2017}]%
        {ruchansky2017csi}
\bibfield{author}{\bibinfo{person}{Natali Ruchansky}, \bibinfo{person}{Sungyong
  Seo}, {and} \bibinfo{person}{Yan Liu}.} \bibinfo{year}{2017}\natexlab{}.
\newblock \showarticletitle{Csi: A hybrid deep model for fake news detection}.
  In \bibinfo{booktitle}{\emph{Proceedings of the 2017 ACM on Conference on
  Information and Knowledge Management}}. \bibinfo{pages}{797--806}.
\newblock


\bibitem[\protect\citeauthoryear{Schlichtkrull, Kipf, Bloem, Van Den~Berg,
  Titov, and Welling}{Schlichtkrull et~al\mbox{.}}{2018}]%
        {schlichtkrull2018modeling}
\bibfield{author}{\bibinfo{person}{Michael Schlichtkrull},
  \bibinfo{person}{Thomas~N Kipf}, \bibinfo{person}{Peter Bloem},
  \bibinfo{person}{Rianne Van Den~Berg}, \bibinfo{person}{Ivan Titov}, {and}
  \bibinfo{person}{Max Welling}.} \bibinfo{year}{2018}\natexlab{}.
\newblock \showarticletitle{Modeling relational data with graph convolutional
  networks}. In \bibinfo{booktitle}{\emph{European semantic web conference}}.
  Springer, \bibinfo{pages}{593--607}.
\newblock


\bibitem[\protect\citeauthoryear{Shu, Mahudeswaran, Wang, Lee, and Liu}{Shu
  et~al\mbox{.}}{2018}]%
        {shu2018fakenewsnet}
\bibfield{author}{\bibinfo{person}{Kai Shu}, \bibinfo{person}{Deepak
  Mahudeswaran}, \bibinfo{person}{Suhang Wang}, \bibinfo{person}{Dongwon Lee},
  {and} \bibinfo{person}{Huan Liu}.} \bibinfo{year}{2018}\natexlab{}.
\newblock \showarticletitle{FakeNewsNet: A Data Repository with News Content,
  Social Context and Dynamic Information for Studying Fake News on Social
  Media}.
\newblock \bibinfo{journal}{\emph{arXiv preprint arXiv:1809.01286}}
  (\bibinfo{year}{2018}).
\newblock


\bibitem[\protect\citeauthoryear{Shu, Wang, and Liu}{Shu
  et~al\mbox{.}}{2019a}]%
        {shu2019beyond}
\bibfield{author}{\bibinfo{person}{Kai Shu}, \bibinfo{person}{Suhang Wang},
  {and} \bibinfo{person}{Huan Liu}.} \bibinfo{year}{2019}\natexlab{a}.
\newblock \showarticletitle{Beyond news contents: The role of social context
  for fake news detection}. In \bibinfo{booktitle}{\emph{Proceedings of the
  twelfth ACM international conference on web search and data mining}}.
  \bibinfo{pages}{312--320}.
\newblock


\bibitem[\protect\citeauthoryear{Shu, Zhou, Wang, Zafarani, and Liu}{Shu
  et~al\mbox{.}}{2019b}]%
        {shu2019role}
\bibfield{author}{\bibinfo{person}{Kai Shu}, \bibinfo{person}{Xinyi Zhou},
  \bibinfo{person}{Suhang Wang}, \bibinfo{person}{Reza Zafarani}, {and}
  \bibinfo{person}{Huan Liu}.} \bibinfo{year}{2019}\natexlab{b}.
\newblock \showarticletitle{The role of user profiles for fake news detection}.
  In \bibinfo{booktitle}{\emph{Proceedings of the 2019 IEEE/ACM international
  conference on advances in social networks analysis and mining}}.
  \bibinfo{pages}{436--439}.
\newblock


\bibitem[\protect\citeauthoryear{Sun, Deng, Nie, and Tang}{Sun
  et~al\mbox{.}}{2019}]%
        {sun2019rotate}
\bibfield{author}{\bibinfo{person}{Zhiqing Sun}, \bibinfo{person}{Zhi-Hong
  Deng}, \bibinfo{person}{Jian-Yun Nie}, {and} \bibinfo{person}{Jian Tang}.}
  \bibinfo{year}{2019}\natexlab{}.
\newblock \showarticletitle{Rotate: Knowledge graph embedding by relational
  rotation in complex space}.
\newblock \bibinfo{journal}{\emph{arXiv preprint arXiv:1902.10197}}
  (\bibinfo{year}{2019}).
\newblock


\bibitem[\protect\citeauthoryear{Veli{\v{c}}kovi{\'c}, Cucurull, Casanova,
  Romero, Lio, and Bengio}{Veli{\v{c}}kovi{\'c} et~al\mbox{.}}{2017}]%
        {velivckovic2017graph}
\bibfield{author}{\bibinfo{person}{Petar Veli{\v{c}}kovi{\'c}},
  \bibinfo{person}{Guillem Cucurull}, \bibinfo{person}{Arantxa Casanova},
  \bibinfo{person}{Adriana Romero}, \bibinfo{person}{Pietro Lio}, {and}
  \bibinfo{person}{Yoshua Bengio}.} \bibinfo{year}{2017}\natexlab{}.
\newblock \showarticletitle{Graph attention networks}.
\newblock \bibinfo{journal}{\emph{arXiv preprint arXiv:1710.10903}}
  (\bibinfo{year}{2017}).
\newblock


\bibitem[\protect\citeauthoryear{Vo and Lee}{Vo and Lee}{2021}]%
        {vo2021hierarchical}
\bibfield{author}{\bibinfo{person}{Nguyen Vo} {and} \bibinfo{person}{Kyumin
  Lee}.} \bibinfo{year}{2021}\natexlab{}.
\newblock \showarticletitle{Hierarchical Multi-head Attentive Network for
  Evidence-aware Fake News Detection}. In \bibinfo{booktitle}{\emph{Proceedings
  of the 16th Conference of the European Chapter of the Association for
  Computational Linguistics: Main Volume}}. \bibinfo{pages}{965--975}.
\newblock


\bibitem[\protect\citeauthoryear{Wang, Ji, Shi, Wang, Ye, Cui, and Yu}{Wang
  et~al\mbox{.}}{2019}]%
        {wang2019heterogeneous}
\bibfield{author}{\bibinfo{person}{Xiao Wang}, \bibinfo{person}{Houye Ji},
  \bibinfo{person}{Chuan Shi}, \bibinfo{person}{Bai Wang},
  \bibinfo{person}{Yanfang Ye}, \bibinfo{person}{Peng Cui}, {and}
  \bibinfo{person}{Philip~S Yu}.} \bibinfo{year}{2019}\natexlab{}.
\newblock \showarticletitle{Heterogeneous graph attention network}. In
  \bibinfo{booktitle}{\emph{The World Wide Web Conference}}.
  \bibinfo{pages}{2022--2032}.
\newblock


\bibitem[\protect\citeauthoryear{Wang, Ma, Jin, Yuan, Xun, Jha, Su, and
  Gao}{Wang et~al\mbox{.}}{2018}]%
        {wang2018eann}
\bibfield{author}{\bibinfo{person}{Yaqing Wang}, \bibinfo{person}{Fenglong Ma},
  \bibinfo{person}{Zhiwei Jin}, \bibinfo{person}{Ye Yuan},
  \bibinfo{person}{Guangxu Xun}, \bibinfo{person}{Kishlay Jha},
  \bibinfo{person}{Lu Su}, {and} \bibinfo{person}{Jing Gao}.}
  \bibinfo{year}{2018}\natexlab{}.
\newblock \showarticletitle{Eann: Event adversarial neural networks for
  multi-modal fake news detection}. In \bibinfo{booktitle}{\emph{Proceedings of
  the 24th acm sigkdd international conference on knowledge discovery \& data
  mining}}. \bibinfo{pages}{849--857}.
\newblock


\bibitem[\protect\citeauthoryear{Wang, Zhang, Feng, and Chen}{Wang
  et~al\mbox{.}}{2014}]%
        {wang2014knowledge}
\bibfield{author}{\bibinfo{person}{Zhen Wang}, \bibinfo{person}{Jianwen Zhang},
  \bibinfo{person}{Jianlin Feng}, {and} \bibinfo{person}{Zheng Chen}.}
  \bibinfo{year}{2014}\natexlab{}.
\newblock \showarticletitle{Knowledge graph embedding by translating on
  hyperplanes}. In \bibinfo{booktitle}{\emph{Proceedings of the AAAI Conference
  on Artificial Intelligence}}, Vol.~\bibinfo{volume}{28}.
\newblock


\bibitem[\protect\citeauthoryear{Wardle and Derakhshan}{Wardle and
  Derakhshan}{2017}]%
        {wardle2017information}
\bibfield{author}{\bibinfo{person}{Claire Wardle} {and}
  \bibinfo{person}{Hossein Derakhshan}.} \bibinfo{year}{2017}\natexlab{}.
\newblock \showarticletitle{Information disorder: Toward an interdisciplinary
  framework for research and policy making}.
\newblock \bibinfo{journal}{\emph{Council of Europe report}}
  \bibinfo{volume}{27} (\bibinfo{year}{2017}), \bibinfo{pages}{1--107}.
\newblock


\bibitem[\protect\citeauthoryear{Zhang, Song, Huang, Swami, and Chawla}{Zhang
  et~al\mbox{.}}{2019}]%
        {zhang2019heterogeneous}
\bibfield{author}{\bibinfo{person}{Chuxu Zhang}, \bibinfo{person}{Dongjin
  Song}, \bibinfo{person}{Chao Huang}, \bibinfo{person}{Ananthram Swami}, {and}
  \bibinfo{person}{Nitesh~V Chawla}.} \bibinfo{year}{2019}\natexlab{}.
\newblock \showarticletitle{Heterogeneous graph neural network}. In
  \bibinfo{booktitle}{\emph{Proceedings of the 25th ACM SIGKDD International
  Conference on Knowledge Discovery \& Data Mining}}.
  \bibinfo{pages}{793--803}.
\newblock


\bibitem[\protect\citeauthoryear{Zhou and Zafarani}{Zhou and Zafarani}{2020}]%
        {zhou2020survey}
\bibfield{author}{\bibinfo{person}{Xinyi Zhou} {and} \bibinfo{person}{Reza
  Zafarani}.} \bibinfo{year}{2020}\natexlab{}.
\newblock \showarticletitle{A survey of fake news: Fundamental theories,
  detection methods, and opportunities}.
\newblock \bibinfo{journal}{\emph{ACM Computing Surveys (CSUR)}}
  \bibinfo{volume}{53}, \bibinfo{number}{5} (\bibinfo{year}{2020}),
  \bibinfo{pages}{1--40}.
\newblock


\end{thebibliography}
\newpage
\appendix

\section{Notations} \label{ap:notations}
\vspace{-10pt}
% The notations used in our paper is summarized as follow:
\begin{table}[h]
    \caption{Notations used in the paper.}
    \ra{1.1}
    \begin{tabular}{ll} 
        \toprule
        \textbf{Notation} & \textbf{Meaning} \\
        \midrule
        $\mathcal{G} = (\mathcal{V},\mathcal{E})$         & Heterogeneous graph of news \\ 
        $\mathcal{V}$   & A set of nodes in the graph $\mathcal{G}$ \\ 
        $\mathcal{E}$   & A set of edges in the graph $\mathcal{G}$ \\
        $\mathcal{R}$   & A set of relations between two nodes (type of edge) \\
        $A$             & A set of types of nodes $A = \{A_p, A_n, A_u\}$\\
        $A_p$           & Node type: publisher \\ 
        $A_n$           & Node type: news \\
        $A_u$           & Node type: user \\
        $\mathcal{P}$   & A set of Meta-Path \\
        $P_U$  & Meta-Path: $News - User - News$ \\
        $P_S$  & Meta-Path: $News - Publisher - News$ \\
        $p$ & A Meta-Path instance \\
        $\mathbf{W}_A$  & Type-specific transformation matrix \\
        $\mathbf{x}_{v}^{A}$ & Initial feature of the node $v$ of type $A$ \\ 
        $\mathbf{h}_{v}^{A}$ & Transformed feature of the node $v$ of type $A$ \\ 
        $f_{enc}$ & Meta-Path instance encoding function \\ 
        $\mathbin\Vert$ & Concatenation operator \\
        $\odot$ & Element-wise product \\
        $\Tilde{\mathbf{v}}$ & Reshape the vector $\mathbf{v}$ in a 2D form \\
        \bottomrule
    \end{tabular}
    \label{t:notations}
    \vspace{-15pt}
\end{table}

\section{Description of Existing Methods}
The bench-marked fake news detection methods can be categorized into text-based approaches and graph-based approaches.
\model{} is also compared with representative graph embedding methods made for the homogeneous and heterogeneous graph.
The detail of the methods we compared is listed below. 

\noindent \textbf{Text-based Methods: }
\begin{itemize}[leftmargin=.15in]
    \item \textbf{TF-IDF + SVM}: TF-IDF is short for term frequency-inverse document frequency. It is intended to represent the importance of a word in a document. Feature vectors were extracted based on news article contents with TF-IDF, and  SVM is applied to it.
    \item \textbf{LIWC}~\cite{pennebaker2015development} \textbf{+ SVM}: LIWC stands for Linguistic Inquiry and Word Count. It is widely used to extract words falling into psychologically meaningful categories, and these words can be used to compose a feature vector.
    \item \textbf{Doc2Vec}~\cite{le2014distributed} \textbf{+ SVM}: Doc2Vec is a paragraph embedding technique based on Word2Vec~\cite{mikolov2013efficient}. It uses skip-gram and CBOW model to learn the representation vector. Doc2Vec is considered as an unsupervised learning approach to learn the latent representation of a document.
\end{itemize}

\noindent \textbf{Graph-based Methods: }
\begin{itemize}[leftmargin=.15in]
        \item \textbf{SAFER}~\cite{chandra2020graph}: SAFER uses GCN and pre-trained RoBERTa model to embed news nodes in the heterogeneous graph. They concatenate two vectors and apply Logistic Regression to classify the news embeddings.
        \item \textbf{CSI}~\cite{ruchansky2017csi}: CSI is a hybrid deep learning based framework that aims to model the response, text, and user engagement of the news. The representation of response and text is concatenated with the user vector and score.
        \item \textbf{FANG}~\cite{nguyen2020fang}: FANG divides the detection task into several sub-tasks, such as textual encoding and stance detection. The final detection object is optimized by defining loss functions for those sub-tasks.
        \item \textbf{AA-HGNN}~\cite{ren2020adversarial}: AA-HGNN uses active learning to tackle the limited training data problem and extends GAT~\cite{velivckovic2017graph} to learn the news representation in the graph.
\end{itemize}

\begin{figure}[t]
\begin{algorithm}[H]
\small
\caption{Learning Algorithm}
\label{alg:train}
\begin{algorithmic}[0]
\Require Heterogeneous Graph of News $\mathcal{G}=(\mathcal{V},\mathcal{E})$,
\Statex node feature $\{\mathbf{x_v}, \forall v \in \mathcal{V} \}$
\Statex node types $\mathcal{A} = \{A_p, A_n, A_u\}$
\Statex label $y$
\Ensure Learn-able parameters $\theta$ % The News embedding $\mathbf{z}_v$

\For{each epoch}
    \For{node type $A \in \mathcal{A}$}
    \State \textit{\textcolor{gray}{\# Node feature transformation}} 
    \State $\mathbf{h}_v^A = \mathbf{W}_A \cdot \mathbf{x}_v^A$
    \EndFor
    
    \For{Meta-Path schema $P \in \mathcal{P}$}
    \State $\mathbf{h}_{p} = f_\theta(\mathbf{h}_u, r, \mathbf{h}_w, r^{-1})$
        \If{$P == \mathcal{P}_S$}
            \State \textit{\textcolor{gray}{\# Calculate the weight coefficient $\alpha_{p}$ for each Meta-Path}}
            \State \textit{\textcolor{gray}{\# instance.}}
            \State $\mathbf{h}_v^P = \mathbin\Vert_{k=1}^{K} \sigma (\sum_{u \in \mathcal{N}_v^P} [\alpha_{p}]_{k} \cdot \mathbf{h}_{p})$
        \EndIf
        \If{$P == \mathcal{P}_U$}
            \State All $\mathbf{h}_{p_i}$ are sorted chronologically
            \State $\mathbf{h}_v^{\mathcal{P}_U} = \textbf{GRU} (\mathbf{h}_{p_1}, \mathbf{h}_{p_2}, ..., \mathbf{h}_{p_n}), p_i \in \mathbf{P_U}$ 
        \EndIf
        \State $s_{P} = \frac{1}{|\mathcal{V}|} \sum_{v \in \mathcal{V}} tanh(\mathbf{M}_A \cdot \mathbf{h}_v^{P} + \mathbf{b}_{A})$
    \EndFor
    \State Calculate the weight coefficient $\beta_{P}$ for each Meta-Path.
    \State $\mathbf{h}_v = \sum_{P \in \mathcal{P}} \beta_P \cdot h_v^P$
    \State \textit{\textcolor{gray}{\# a fully connected layer for new classification.}} 
    \State $\mathbf{z}_v = W_c \cdot \mathbf{h}_v$ 
    % \State \textit{\textcolor{gray}{\# Calculate the probability of fake and real news}}
    % \State \textit{\textcolor{gray}{\# according to the output logits, $\mathbf{z}_v$}}
    \State $[\mathbf{P}_{real}, \mathbf{P}_{fake}] = softmax(\mathbf{z}_v)$
    \State $\mathcal{L} = -\sum(ylog(\mathbf{P}_{fake})) + (1-y)log(\mathbf{P}_{real})$
    \State $\theta \gets Backpropagate(\mathcal{L})$
\EndFor
\end{algorithmic}
\end{algorithm}
\vspace{-35pt}
\end{figure}

\noindent \textbf{GNN baselines}: 
\begin{itemize}[leftmargin=.15in]
    \item \textbf{GCN}~\cite{kipf2016semi}: GCN is a deep learning based method on a graph-structured data. Each node is learned by aggregating the feature information from its neighbors and the feature of itself. 
    \item \textbf{GAT}~\cite{velivckovic2017graph}: GAT is similar to GCN, but it introduces the attention mechanism to replace the statically normalized convolution operation in GCN.
    \item \textbf{GraphSAGE}~\cite{hamilton2017inductive}: GraphSAGE is a general inductive framework that learns a node representation by sampling its neighbors and aggregating features of sampled nodes. 
    \item \textbf{R-GCN}~\cite{schlichtkrull2018modeling}: R-GCN is an application of the GCN framework for modeling relational data. In R-GCN, edges can represent different relations.
    \item \textbf{HAN}~\cite{wang2019heterogeneous}: HAN is an extension of GAT on the heterogeneous graph. Meta-Path extraction strategy and attention mechanism are adopted to learn the representation of a node.
\end{itemize}

% \section{Learning Algorithm}

\section{Validation Loss during Training}
\begin{figure}[h]
\begin{subfigure}{0.49\columnwidth}
  \centering
  \raisebox{-\height}{\includegraphics[width=\textwidth]{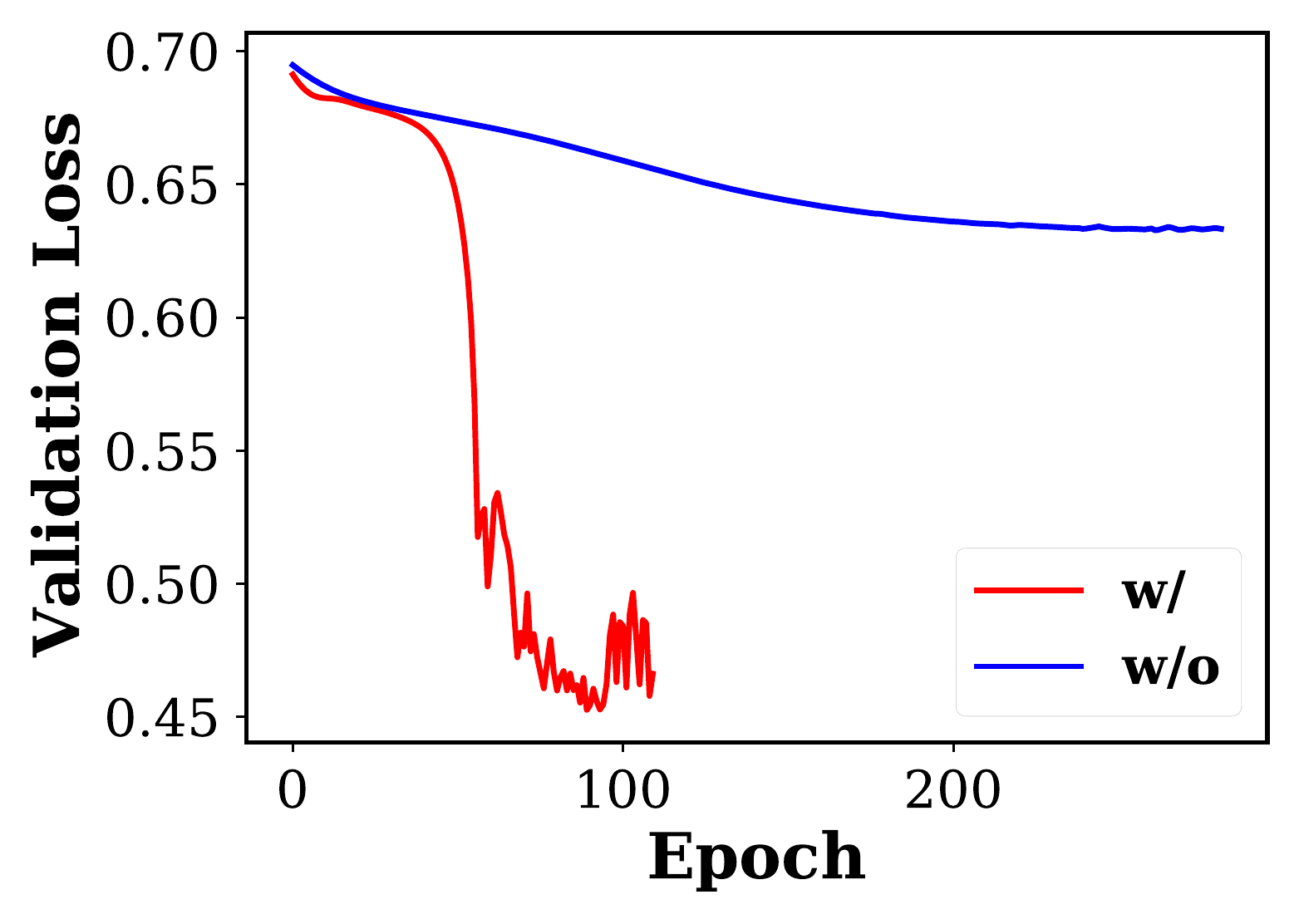}}
  \caption{FANG}
  \label{fig:mis_sub-first}
\end{subfigure}
\begin{subfigure}{0.49\columnwidth}
  \centering
  \raisebox{-\height}{\includegraphics[width=\textwidth]{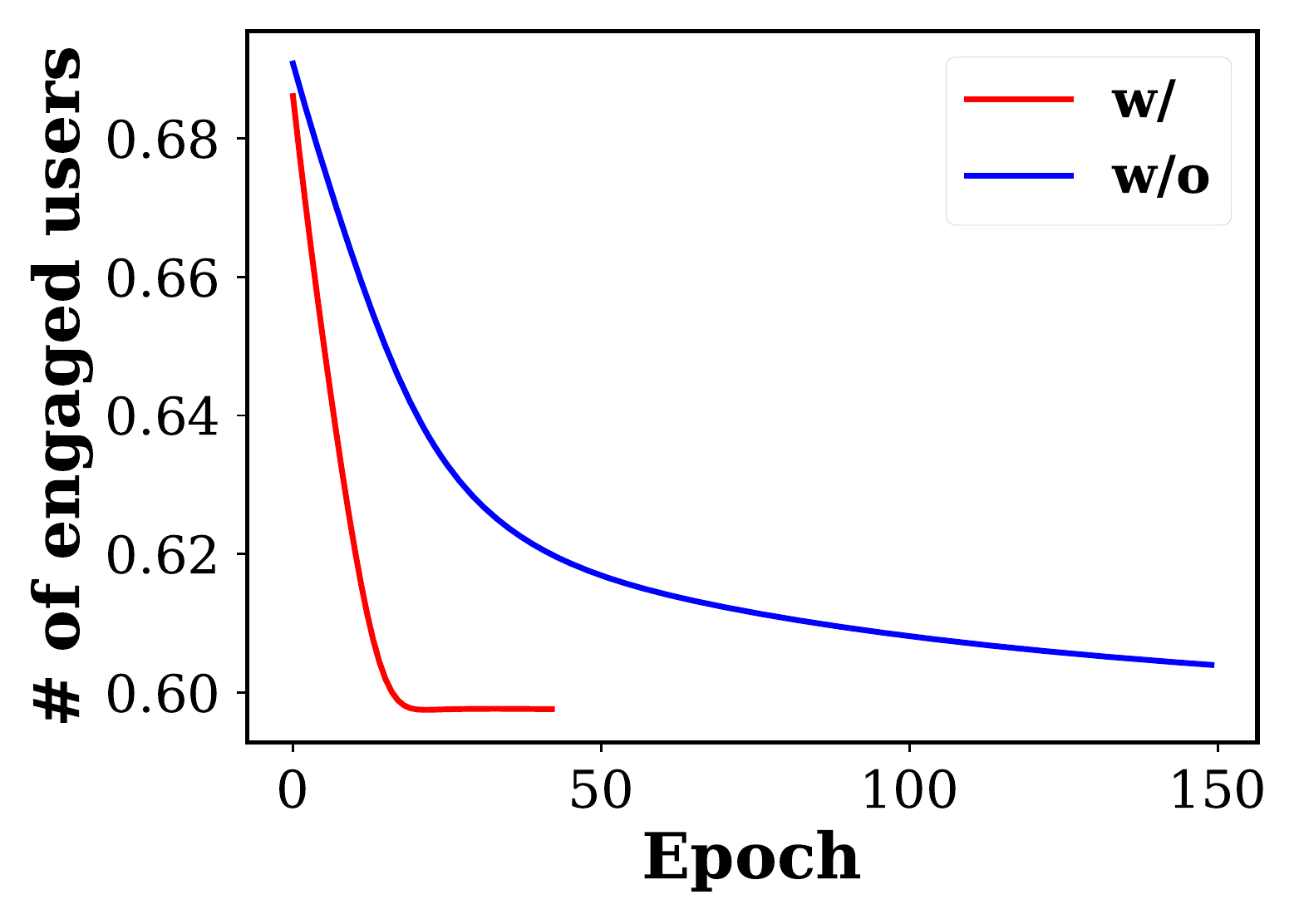}}
  \caption{HealthStory}
  \label{fig:mis_sub-second}
\end{subfigure}
\caption{Validation loss during training. (Red line indicates the validation loss of \model{} with temporal information, blue line indicates the validation loss of \model{} without temporal information.)}
\label{fig:mis_dis_val}
\vspace{-10pt}
\end{figure}

In Section~\ref{sec:mis_dis}, to see the impact of temporal information in \model{}, we replace the RNN with attention mechanism.
In other words, we compare the \model{} trained with and without temporal information.
These two \model{} are trained with two dataset, and corresponding validation loss during the training is shown in Figure~\ref{fig:mis_dis_val}.
The \model{} trained with temporal information has faster convergence speed than \model{} trained without temporal information in FANG dataset; In the HealthStory dataset, however, two models have no significant difference.
Considering that fake news in FANG dataset is disinformation, and fake news in HealthStory is misinformation, temporal information can accelerates the convergence speed of training when identifying disinformation.

\section{Ablation study on Meta-Path Instance Encoding Methods}

In Section~\ref{method:encoding}, we propose to use knowledge triple embedding methods to encode Meta-Path instances, and we adopt TransE in \model{}.
We wanted to examine the performance differences by changing the Meta-Path encoding method to other knowledge triple embedding methods, RotatE and ConvE.
Descriptions of the three encoding methods are introduced below.

\begin{itemize}[leftmargin=.15in]
    \item \textbf{TransE}~\cite{wang2014knowledge}: The TransE model represents relations as translations and aims to model the inversion and composition patterns. It defines each relation as a translation from the subject entity to the object entity.
    \item \textbf{RotatE}~\cite{sun2019rotate}: The RotatE model maps the entities and relations to the complex vector space and defines each relation as a rotation from the subject entity to the object entity.
    \item \textbf{ConvE}~\cite{dettmers2018convolutional}: The ConvE model uses 2D convolution over embedding and multiple layers of nonlinear features to mode knowledge graphs. They reshape the embedding of subject and predicates in a 2D form and apply convolution calculations on it. 
\end{itemize}

To show the performance differences when different knowledge triple embedding methods are applied, F1 score, Accuracy, and AUC score were measured on two datasets: FANG and HealthStory.
Table~\ref{t:vs_other_encoder} indicates that TransE gives better results than the others. 
This reason can be drawn from the fact that TransE requires fewer parameters and operations than RotatE and ConvE.
With limited training data, complex models are easy to suffer from over-fitting, which will cause performance degradation.

\begin{table}[h]\centering
    \caption{Performance of detection result when apply different Meta-Path encoding method. Bold texts indicate the best encoding method in \model{}.}
    \vspace{5pt}
    \ra{1.2}
	\begin{tabular}{cccc}
	\toprule
	% & \multicolumn{3}{c}{\textbf{FANG}} & \\ % \phantom{abc} & \multicolumn{2}{c}{\textbf{HealthStory}} \\
	% \cline{2-4}
	& \textbf{F1 Score} & \textbf{Accuracy} & \textbf{AUC} \\
	\hline
    \textbf{TransE} & \textbf{0.831$\pm$0.037} & \textbf{0.831$\pm$0.037} & \textbf{0.900$\pm$0.057} \\ % & 0.532 & 0.568 \\
    \textbf{RotatE} & 0.799$\pm$0.035 & 0.799$\pm$0.036 & 0.862$\pm$0.035 \\ % & 0.526 & 0.513 \\
    \textbf{ConvE} & 0.532$\pm$0.174 & 0.526$\pm$0.079 & 0.665$\pm$0.021 \\ % & 0.693 & 0.665 \\
    \bottomrule
    \end{tabular}
    \label{t:vs_other_encoder}
    % \vspace{-0.3cm}
\end{table}

\section{t-SNE visualization} \label{ap:tsne}
\begin{figure}[ht]
\captionsetup{justification=centering}
\begin{subfigure}{0.49\columnwidth}
  \centering
    \raisebox{-\height}{\includegraphics[width=\textwidth]{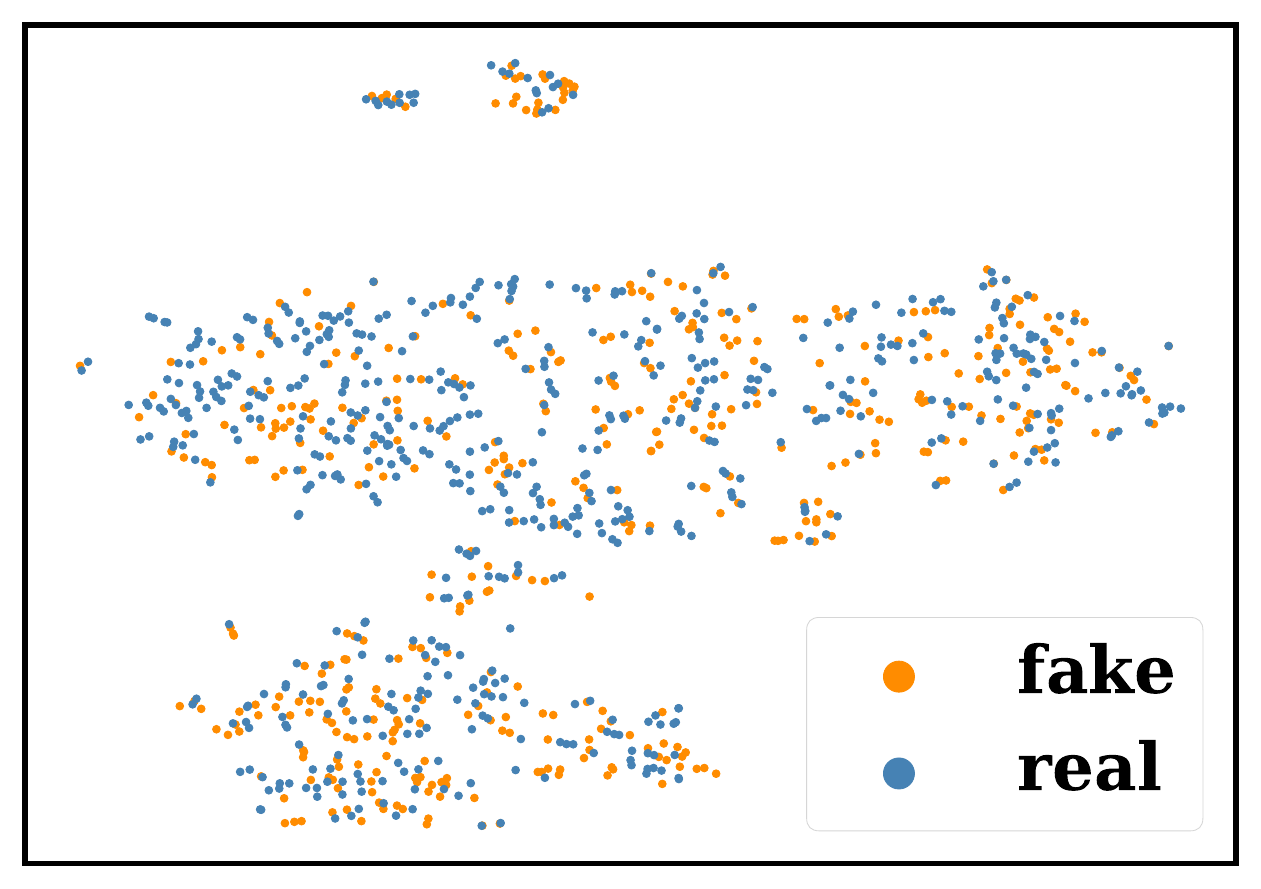}}
  \caption{FANG}
  \label{fig:tsne_fang}
\end{subfigure}
\begin{subfigure}{0.49\columnwidth}
  \centering
    \raisebox{-\height}{\includegraphics[width=\textwidth]{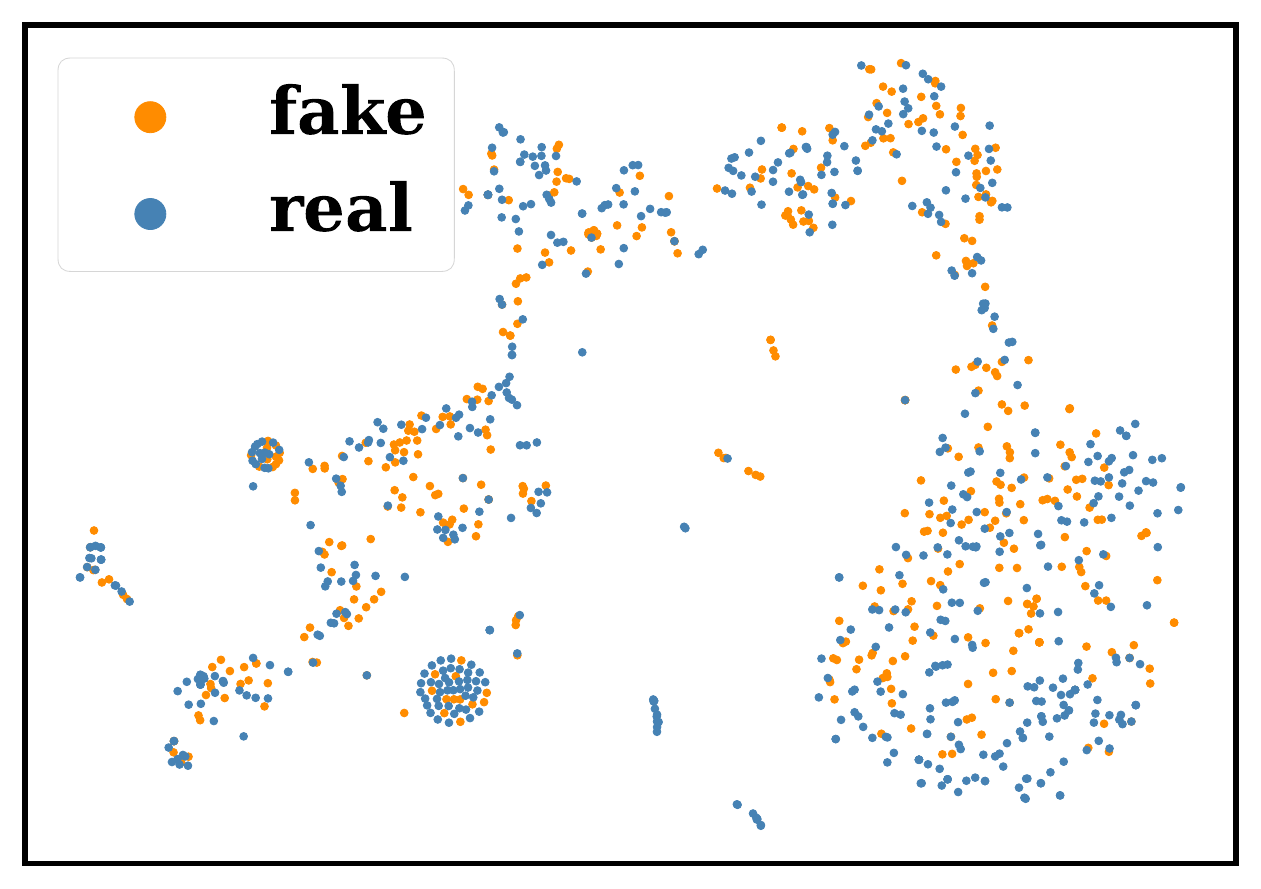}}
  \caption{SAFER}
  \label{fig:tsne_safer}
\end{subfigure}
\begin{subfigure}{0.49\columnwidth}
  \centering
  \raisebox{-\height}{\includegraphics[width=\textwidth]{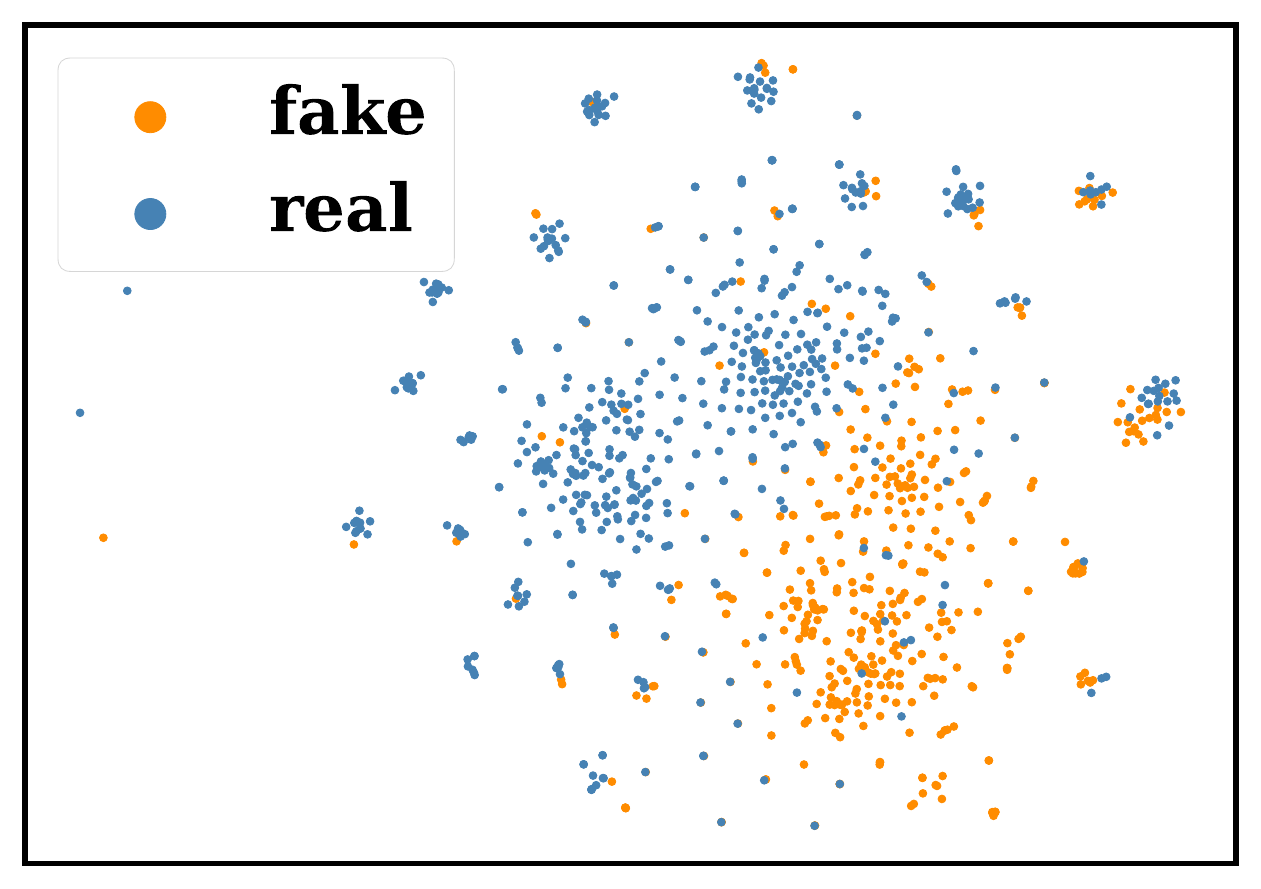}}
  \caption{\model{} \\ (w/o temporal information)}
  \label{fig:tsne_attn}
\end{subfigure}
\begin{subfigure}{0.49\columnwidth}
  \centering
  \raisebox{-\height}{\includegraphics[width=\textwidth]{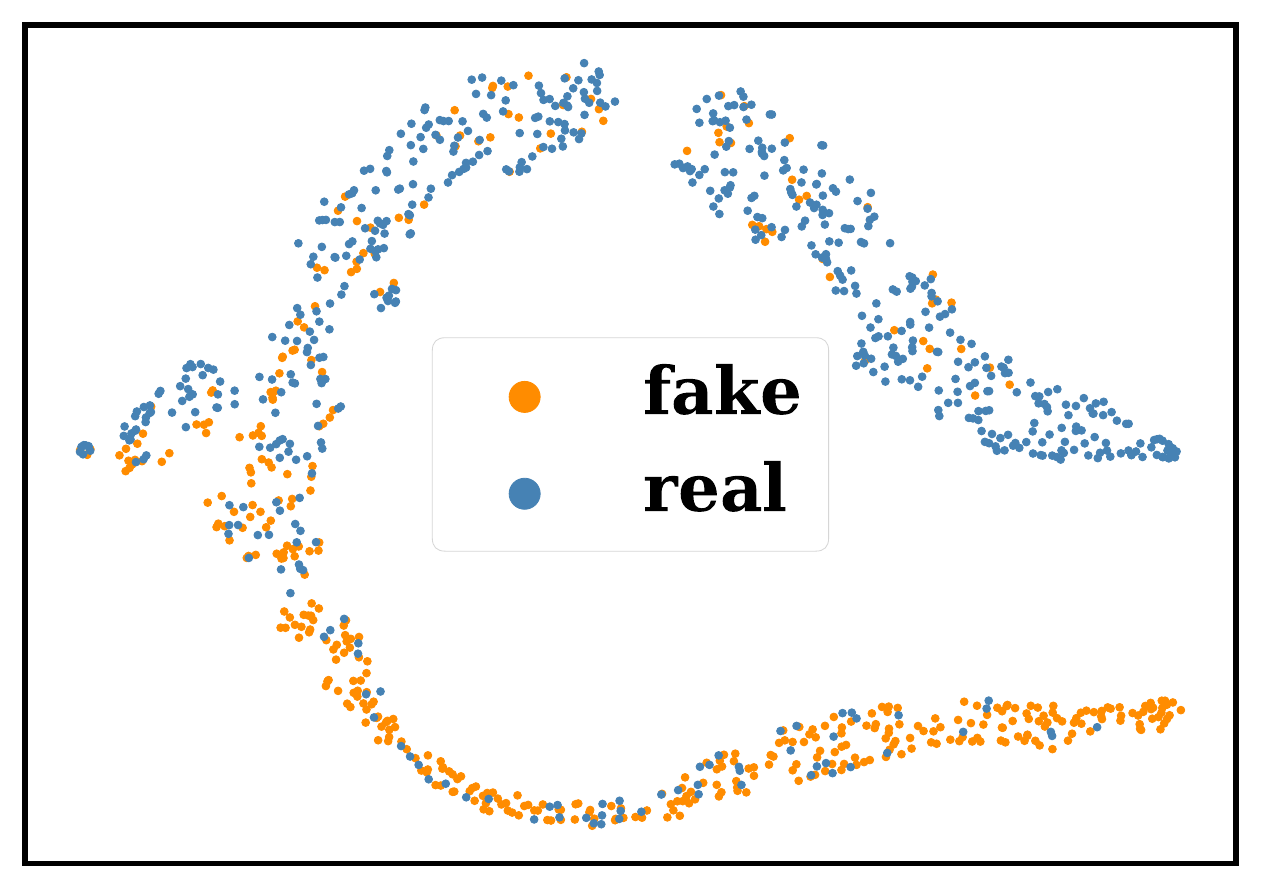}}
  \centering
  \caption{\model{} \\ (w/ temporal information)}
  \label{fig:tsne_rnn}
\end{subfigure}
\caption{t-SNE visualization of news representations.}
\label{fig:tsne}
\end{figure}

To show that the news representation produced by \model{} is better than the existing methods, t-SNE was adopted to visualize news representation in a two-dimensional plane (Figure~\ref{fig:tsne}). 
The t-SNE technique is a well-known method to visualize the high-dimensional data in a two-dimensional plane~\cite{maaten2008visualizing}. 
As can be seen in Figure~\ref{fig:tsne}, the representations of \model{} are clustered tighter than the other methods, implying a significant improvement over existing methods.

\newpage

\end{document}